\documentclass[a4paper,fleqn,usenatbib]{mnras}

\usepackage{newtxtext,newtxmath}

\usepackage[T1]{fontenc}\usepackage{ae,aecompl}

\usepackage{graphicx}	
\usepackage{amsmath}	
\usepackage{amssymb}	

\title[Kinematics of isolated early-type dwarfs]{
Implications for the origin of early-type dwarf galaxies --- \\ the discovery of rotation in isolated, low-mass early-type galaxies}

\author[J. Janz et al.]{
Joachim Janz,$^{1}$\thanks{Email: joachim.janz@gmail.com} Samantha J. Penny,$^2$ Alister W. Graham,$^1$ Duncan A. Forbes,$^1$ \newauthor
Roger L. Davies$^{3}$\\
$^1$Centre for Astrophysics and Supercomputing, Swinburne University, Hawthorn, VIC 3122, Australia\\
$^2$Institute of Cosmology and Gravitation, University of Portsmouth, Dennis Sciama Building, Burnaby Road, Portsmouth PO1 3FX, UK\\
$^3$Sub-department of Astrophysics, Department of Physics, University of Oxford, Denys Wilkinson Building, Keble Road, Oxford OX1 3RH, UK}
\date{Accepted 2017 March 10. Received 2017 March 2; in original form 2016 October 7}
\pubyear{2017}

\begin{document}
\label{firstpage}
\pagerange{\pageref{firstpage}--\pageref{lastpage}}
\maketitle

\begin{abstract}
We present the discovery of {rotation} in quenched, low-mass early-type galaxies that are isolated.
This finding challenges the claim that (all) rotating dwarf early-type galaxies in clusters were once 
spiral galaxies that have since been harassed and transformed into  early-type galaxies.
Our search of the Sloan Digital Sky Survey data within the Local volume ($z<0.02$) has
yielded a sample of 46 galaxies with a stellar mass $M_\star \lesssim 5\times10^9\, \textrm{M}_\odot$ { (median $M_\star \sim 9.29 \times 10^8\, \textrm{M}_\odot$)}, 
a low H$\alpha$ equivalent width EW$_{{\rm H}\alpha}< 2$~\AA{},
and no massive neighbour ($M_{\star}\gtrsim3 \times 10^{10}$~M$_{\odot}$) within a velocity interval of $\Delta V = 500$ km~s$^{-1}$ and a projected distance 
of $\sim1$ Mpc.
Nine of these galaxies  were subsequently observed with Keck ESI and their radial kinematics are presented here.
These extend out to  the half-light radius $R_\textrm{e}$ in the best cases, and beyond  $R_\textrm{e}/2$ for all.
They reveal a variety of behaviours similar to those of a comparison sample of  early-type dwarf galaxies
in the Virgo cluster observed by Toloba et al. Both samples have similar frequencies of slow and fast rotators, as well as kinematically
decoupled cores. This, and especially the finding of rotating quenched low-mass galaxies  in isolation,
reveals that the early-type dwarfs in galaxy clusters need not be harassed or tidally stirred spiral galaxies.
\end{abstract}




\section{Introduction}
 { Early-type dwarf galaxies (dE/dS0s) are the dominant galaxy type in galaxy clusters but are rare in the field \citep{1988ARA&A..26..509B}.
The prevalent conclusion is that a high density environment is essential for their formation.
Early-type dwarfs are characterised by $B$-band luminosity fainter than $M_B > -18$~mag ($M_r > -19.3$~mag),
corresponding to stellar masses below $M_{\star} < 5\times10^{9}$~M$_{\odot}$, early-type morphology,
and a predominantly quenched stellar population.
Some of them exhibit features usually connected to discs, such as spiral arms and bars, as well as other morphological signs of a 
disc \citep{binggeli_cameron,Jerjen:2000tm,2002A&A...391..823B,deRijcke:2003jq,2003AJ....126.1794G,2003AJ....125.2936G,2006AJ....132..497L,2014MNRAS.443.3381P}.
 Their average shape is typical of  a thick disc \citep{2010MNRAS.406L..65S}. 
 Often these features are thought to be inherited from a disc-like progenitor.
 While it is debated, whether  structural decompositions can be used to support or constrain such a transformation scenario \citep{2005AJ....130..475A,2012ApJ...745L..24J,2014ApJ...786..105J,2016A&A...587A.111A}, the significant rotational component  found in the kinematics 
 of several early-type dwarfs (e.g.~\citealt{Pedraz:2002ba,2003AJ....126.1794G}, and many others since then)  is sometimes understood 
 as direct evidence for them being the remnants of spiral or irregular disc galaxies, which were transformed by
 the high density environment.
 
 Physical processes that are thought to explain such transformations via gas removal and dynamical heating unavoidably
 act in these environments due to the presence of hot gas, a high density of galaxies, and deep potential wells
 (i.e.~ram pressure stripping and harassment; \citealt{1972ApJ...176....1G,Moore:1996il}).
 Recently, a large fraction of dwarf galaxies in the Virgo cluster were shown to be rotating \citep{2015ApJ...799..172T}. 
Similarly,
\citet{2016MNRAS.462.3955P}  found that the majority of quenched dwarfs in the first year of data from the 
MaNGA survey are also rotating, and many have disc-like morphologies.
 This raises the question whether all early-type dwarfs in galaxy clusters are transformed discs.

While early-type dwarfs in the field are rare, they do still exist \citep[e.g.][]{2006AJ....131..806G}.
The finding that such galaxies rotate will invalidate the conclusion that rotating early-type dwarfs 
necessarily were spiral galaxies that were then transformed by the cluster environment.
In fact, for the early-type dwarfs in clusters it has been suggested that only the combination of several different processes 
can   explain the varied properties of early-type dwarfs \citep[see, e.g.,][for an overview]{2009AN....330.1043L}.
These include the aforementioned ram pressure stripping and harassment, but also starvation and tidal stirring \citep[e.g.][]{2008ApJ...674..742B,2001ApJ...547L.123M}.

Another alternative  \citep[e.g.][]{deRijcke:2005hq,2008ApJ...689L..25J,2009ApJ...696L.102J}  to the harassment of (more massive) disc galaxies
is particularly relevant when considering the quenched low-mass objects in low density environments or isolation: 
 In the $\Lambda$CDM model of the Universe, low-mass dwarf galaxies are the first galaxies to form. These semi-pressure supported systems are the building blocks of
 massive galaxies in a hierarchically assembling Universe.  
 They may have preferentially formed in over-dense regions that became galaxy groups and clusters.  
 
More massive early-type galaxies often contain a significant rotating disc (e.g., ATLAS$^\textrm{3D}$, \citealt{Emsellem:2011br,2014MNRAS.441..274S}, and references therein; SLUGGS, \citealt{2014ApJ...791...80A,2016MNRAS.457..147F}). Nonetheless, not all of these galaxies are
 thought to be transformed spirals. This already indicates that the presence of a rotating disc in early-type galaxies
 is not necessarily conclusive for a (more massive) spiral progenitor.
 Like these higher mass early-type galaxies, some early-type dwarfs too may have acquired
  a disc around a pre-existing, possibly pressure supported, (lower mass) progenitor (e.g., \citealt{2015ApJ...804...32G}, 
and references therein),  may have continuously grown from a fainter disc \citep{2016arXiv160805412T}, or may just have formed in a dissipate collapse  \citep[e.g.][]{2014MNRAS.444.3357N} to become dwarf lenticular (dS0) galaxies. }

In this study, we present a systematic search for isolated quenched low-mass galaxies (Section~\ref{section:sample}), which allowed us to select suitable candidates
for follow-up spectroscopy (Section~\ref{section:data}). The data analysis is described in Section~\ref{section:analysis}, and in Section~\ref{section:results} we
characterise the kinematics of our objects and compare them to those of early-type dwarfs in the Virgo cluster.
Possible ways to form isolated galaxies like those in our sample are discussed in Section~\ref{section:discussion}.
Moreover, the important implications of our findings of rotating  quenched low-mass galaxies in isolation
for the formation scenarios of early-type dwarfs in clusters are highlighted in this Section.

\section{Sample selection}
\label{section:sample}

{ We seek to identify low-mass galaxies in the Local Universe that are  quenched and located in very low density environments. 
\citet{2012ApJ...757...85G} searched for isolated quenched galaxies and found none below a stellar mass of $M_\star<10^9$ M$_\odot$, but more than 300 with a stellar mass of  $10^9$ M$_\odot < M_\star < 10^{10}$ M$_\odot$. Following their methodology, we search for  quenched galaxies in the mass regime of dwarf galaxies without a close bright neighbour. The data and our selection criteria are described below.}

The large data volume of the Sloan Digital Sky Survey \citep[SDSS,][]{2011AJ....142...72E} spectroscopic sample is utilised and queried  via the NASA Sloan Atlas.\footnote{\url{http://www.nsatlas.org}} Distances are calculated based on the recession velocities and assuming a Hubble flow 
(throughout  our analysis we use $H_0=70$ km~s$^{-1}$ Mpc$^{-1}$, $\Omega_m=0.3$, $\Omega_\Lambda=0.7$). Our search covers $z<0.02$, comparable to the distance of the Coma cluster ($z=0.0231$; $D=100$~Mpc).
At the SDSS spectroscopic limit $r_{\rm Petrosian} = 17.77$~mag for galaxies, all targets at $z=0.02$ with spectroscopic redshifts will be brighter than $M_{r} = -17.11$~mag.
Generally the completeness will vary as function of the position in sky and local galaxy density, but is better for less crowded fields. 
To restrict the sample to low-mass galaxies, an upper stellar mass cut  of $M_\star <2.45\times10^{9}$~M$_{\odot}$ as listed in the NASA Sloan Atlas was applied, corresponding to $M_\star < 5\times10^{9}$~M$_{\odot}$ using $H_0=70$ km~s$^{-1}$ Mpc$^{-1}$ (the NASA Sloan Atlas employed  $H_0=100$ km~s$^{-1}$ Mpc$^{-1}$). The central velocity dispersions  were required to be smaller than $\sigma < 100$~km~s$^{-1}$, consistent with dwarf galaxies previously examined in the literature.
This selects  LMC mass and lower (see \citealt{2016MNRAS.462.3955P}). All candidates in the final sample  have $B$-band magnitudes of $M_B\gtrsim-18$~mag, the classical demarcation line between normal and dwarf galaxies { \citep[e.g.][]{ferguson_binggeli}.}

\begin{figure}  
   \centering
   \includegraphics[width=0.45\textwidth,angle=0]{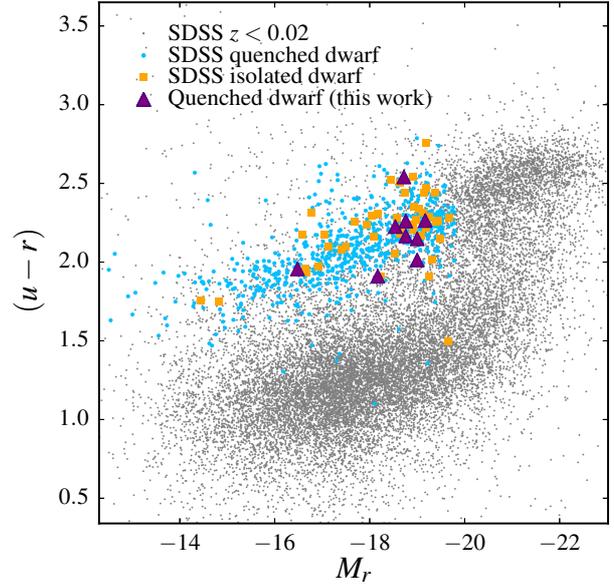} 
       \caption{Colour-magnitude diagram for galaxies in the NASA Sloan Atlas with $z<0.02$ ($D\sim86$\,Mpc). Objects with lower redshift ($z < 0.005$) have been excluded from the plot. Quenched dwarf galaxies in all environments with $M_{B} > -18$~mag are shown as blue points, those dwarfs with a projected separation $D_{\rm proj} >1$~Mpc from their nearest bright neighbour { within a velocity interval of 500 km s$^{-1}$} are shown as orange squares. Our Keck ESI targets are shown as purple triangles. All targets fall  on the galaxy red sequence.}
       \label{fig:psample}
\end{figure}

To identify quenched objects we followed the prescription of \citet{2012ApJ...757...85G}, selecting galaxies with a strong 4000\,\AA{} break, D$_{n}(4000) > 0.6+0.1\times \mathrm{log_{10}}(M_{\star}/\textrm{M}_{\odot})$, and little or no H$\alpha$ emission, with equivalent widths $\mathrm{EW_{H\alpha}} <2$~\AA{}. 
This requirement is met by 665 galaxies with a redshift of  $z< 0.02$, which lie on the red sequence (Fig.~\ref{fig:psample}).
We note that we did not apply any morphological criterion, but this procedure effectively removed  late-type, dwarf irregular galaxies. 
However, our selection criteria exclude a number of early-type dwarfs with blue cores, e.g.~objects like those in the Virgo cluster classified as dE(bc)  by \citet{2006AJ....132.2432L}.
The NASA Sloan Atlas contains 31 of their 38 dE(bc)s, and 16 of them were excluded by our sample selection since they have $\mathrm{EW_{H\alpha}} >2$~\AA{}. 
We note that this potentially excludes a higher fraction of early-type dwarfs in less dense environments, since these typically have blue cores and residual star formation \citep[e.g.][]{2008AJ....135.1488T,2014MNRAS.445..630P}.
{ LEDA 2108986 (Graham et al.~2016) was initially excluded at this stage, since it did not meet the limit in H$\alpha$ equivalent width  (EW$_{{\rm H}\alpha} > 2$~\AA{}).
We include this galaxy here, since it is well isolated ($D_{\rm proj} = 3.33$~Mpc), has a strong 4000\,\AA{} break $D_{n}(4000) = 1.24$~\AA{},
is rather faint  ($M_{B} = -16.96$~mag), and has early-type morphology.}
Galaxies with clear signs of spiral structures, an AGN, and star formation in the outer regions (which was missed by the SDSS fibre), along with any satellite trails, were manually removed after a visual inspection of their SDSS colour imaging. 
All galaxies finally selected for spectroscopy are of smooth appearance devoid of irregularities and mostly quite featureless.

\begin{figure}  
   \centering
   \includegraphics[width=0.45\textwidth,angle=0]{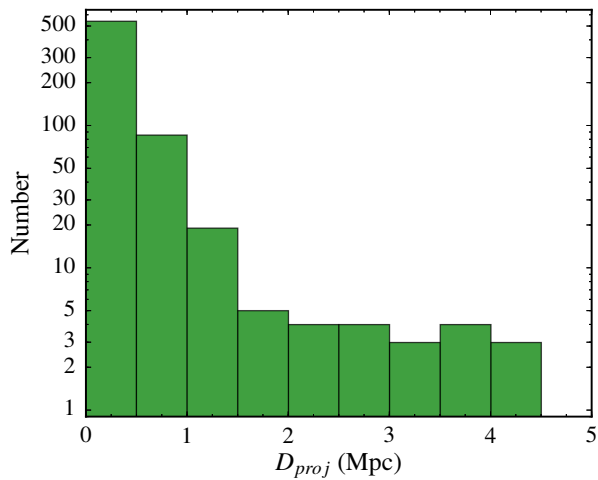} 
      \caption{Histogram showing the projected separation of each quenched dwarf in the NASA Sloan Atlas with $z< 0.02$ (blue points in Fig.~\ref{fig:psample}) from a luminous galaxy with $M_{K_s}$<$-23$~mag ($M_{\star}\gtrsim3 \times 10^{10}$~M$_{\odot}$) within a velocity interval of $\pm$$500$~km~s$^{-1}$. The majority of dwarfs are located at small separations from a nearby luminous galaxy, with a mean projected separation $D_{\rm proj}= 0.36$~Mpc. }
   \label{fig:dhist}
\end{figure}

\begin{figure*}  
   \centering
   \includegraphics[width=0.33\textwidth,angle=270]{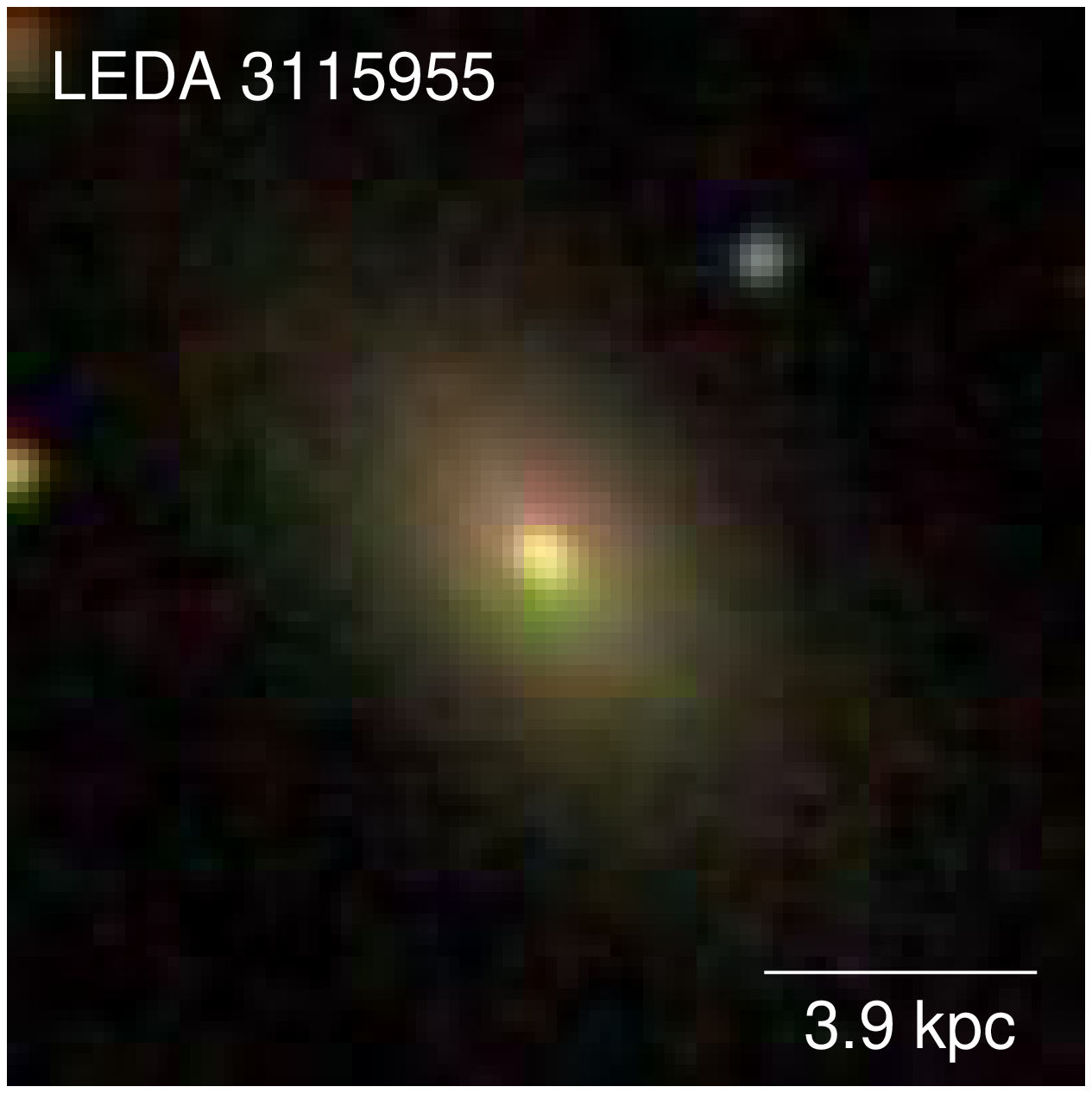} 
   \includegraphics[width=0.33\textwidth,angle=270]{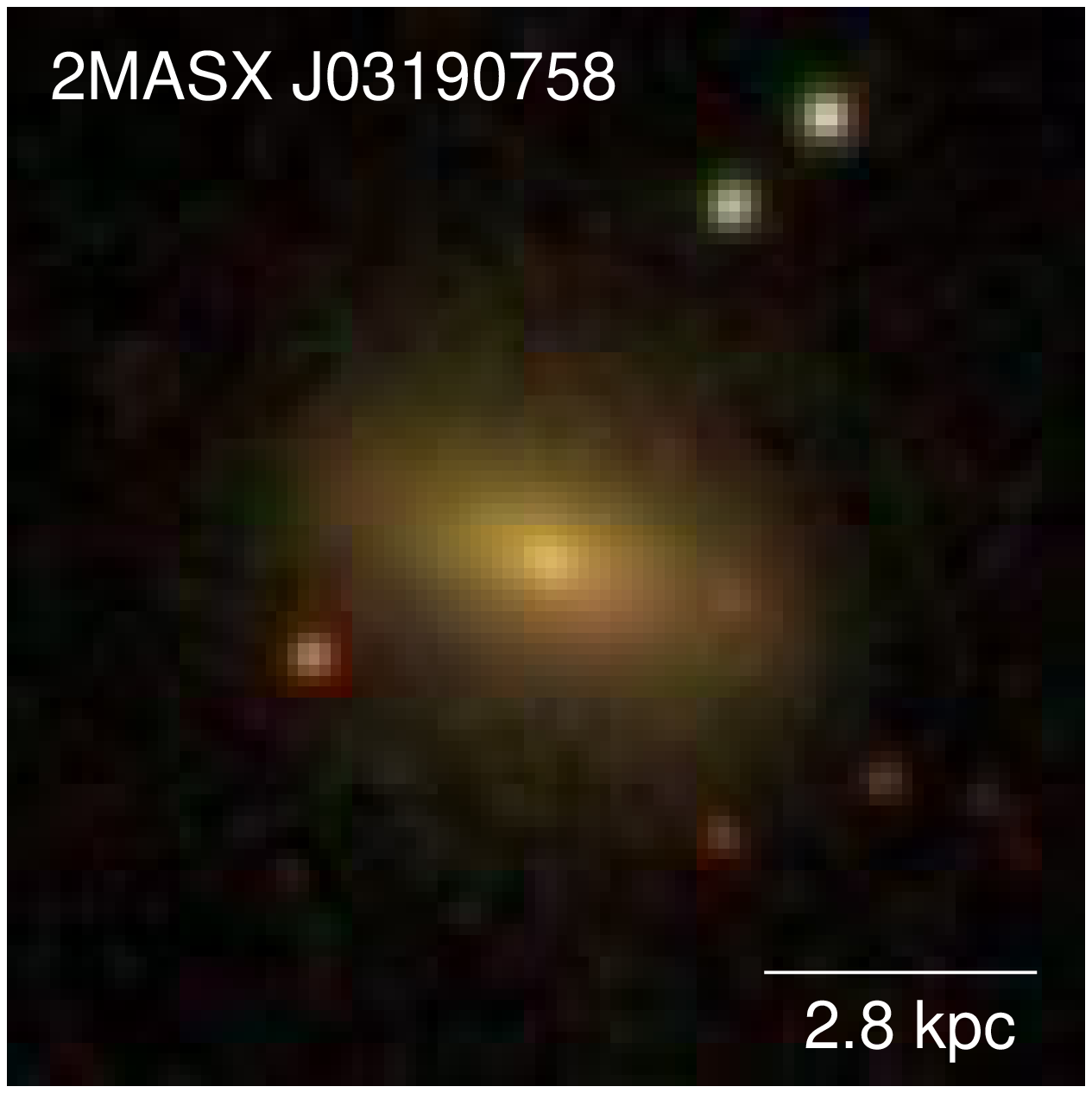} 
   \includegraphics[width=0.33\textwidth,angle=270]{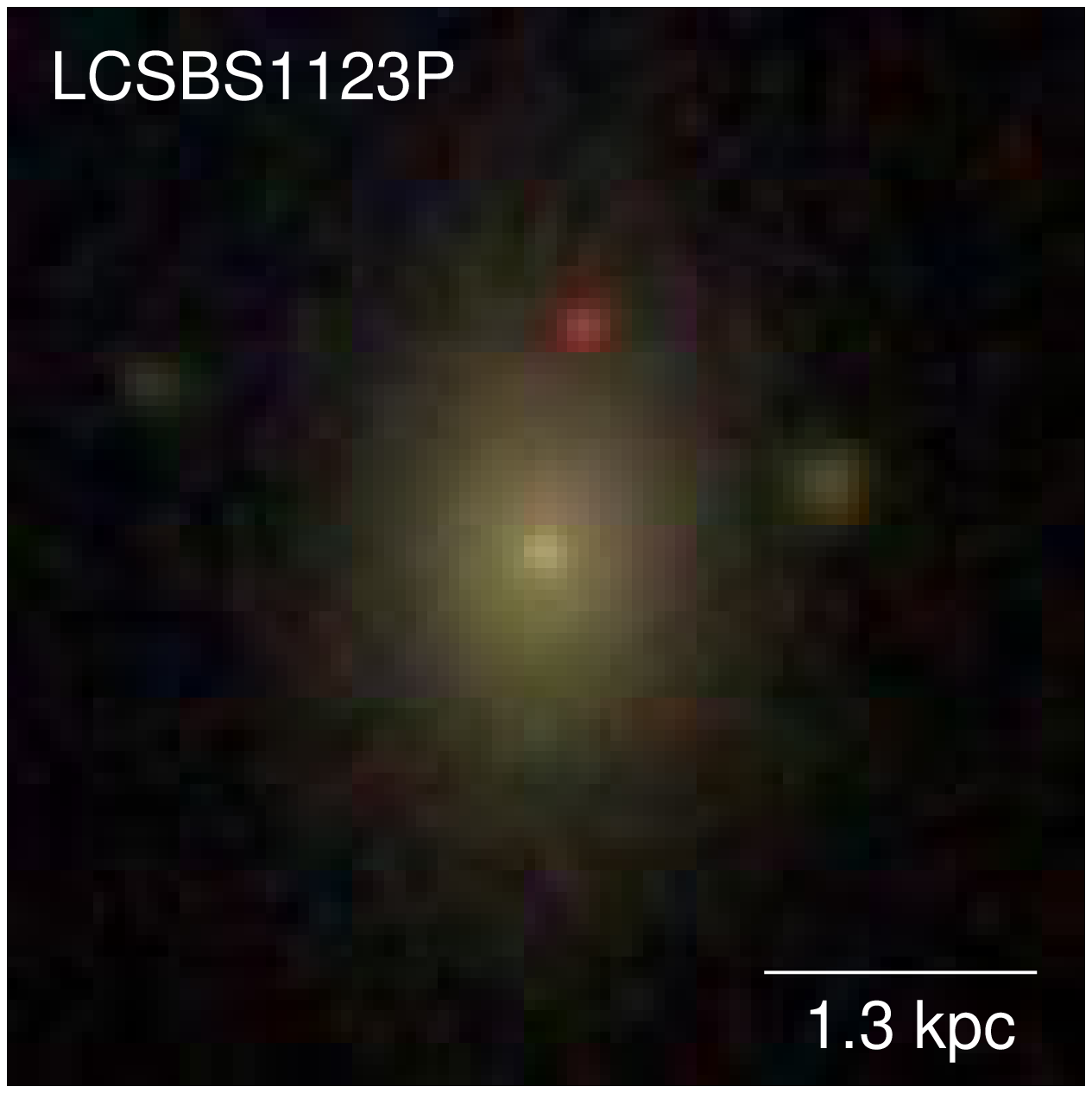} \\
   \includegraphics[width=0.33\textwidth,angle=270]{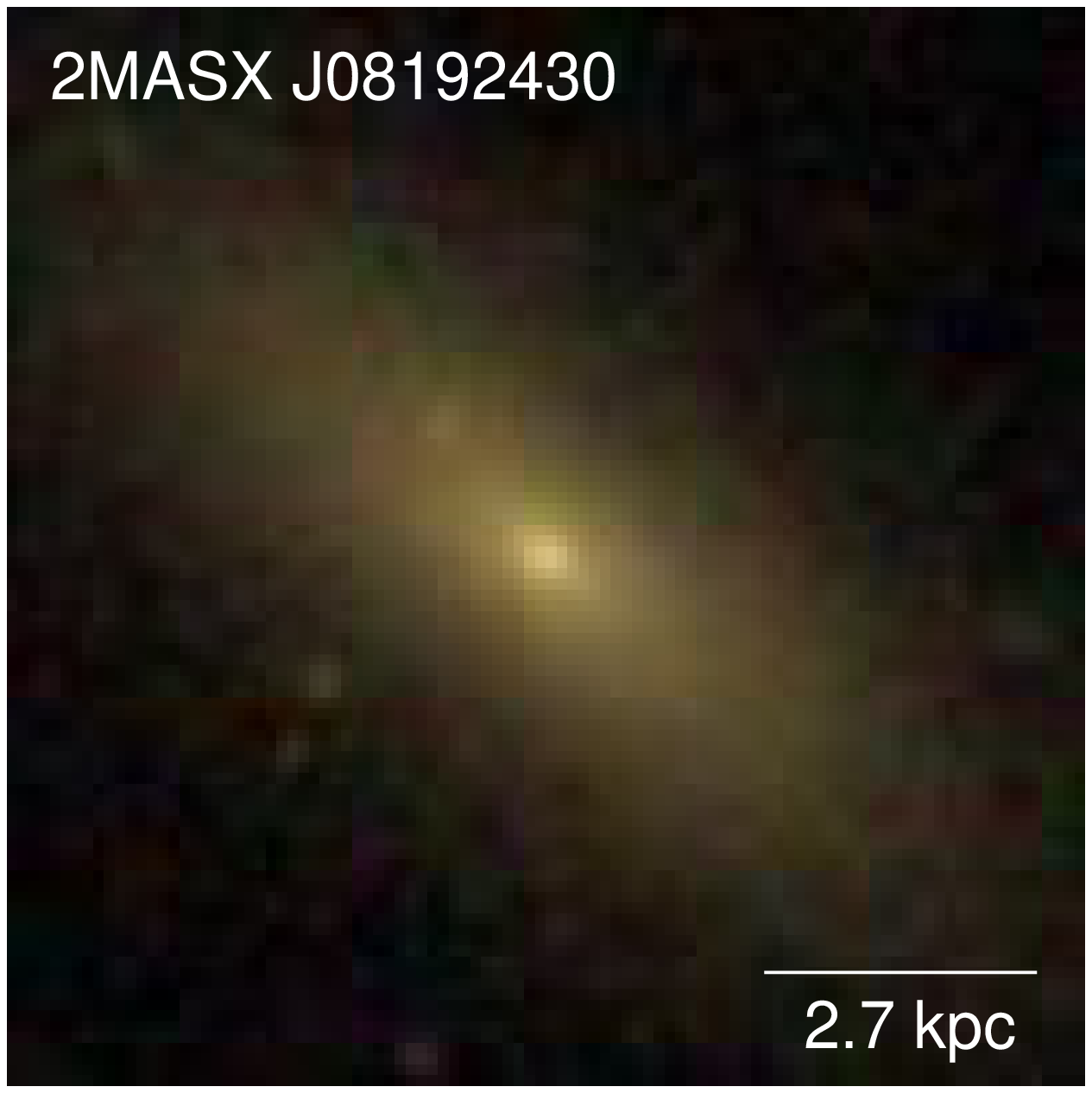} 
   \includegraphics[width=0.33\textwidth,angle=270]{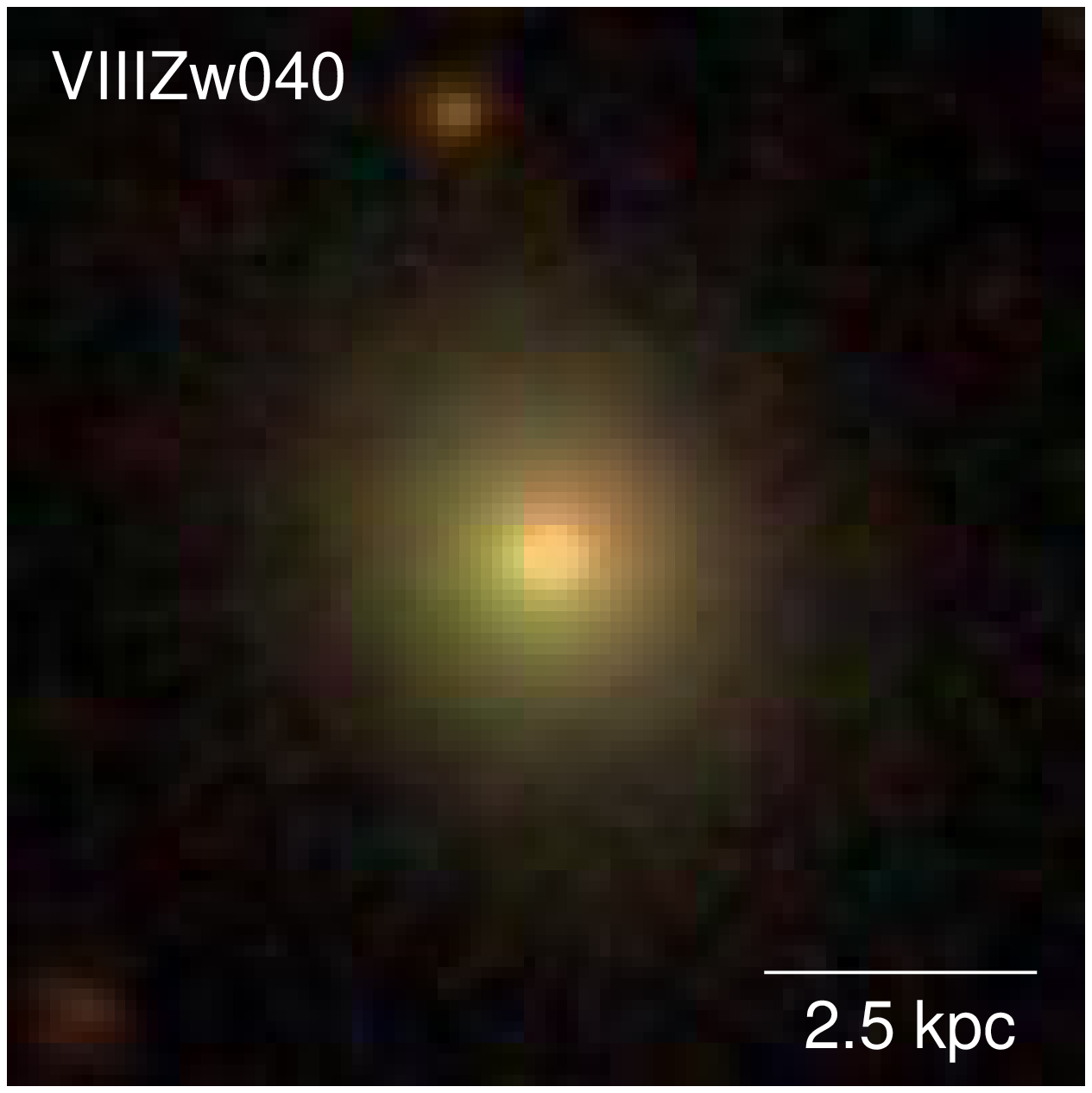} 
   \includegraphics[width=0.33\textwidth,angle=270]{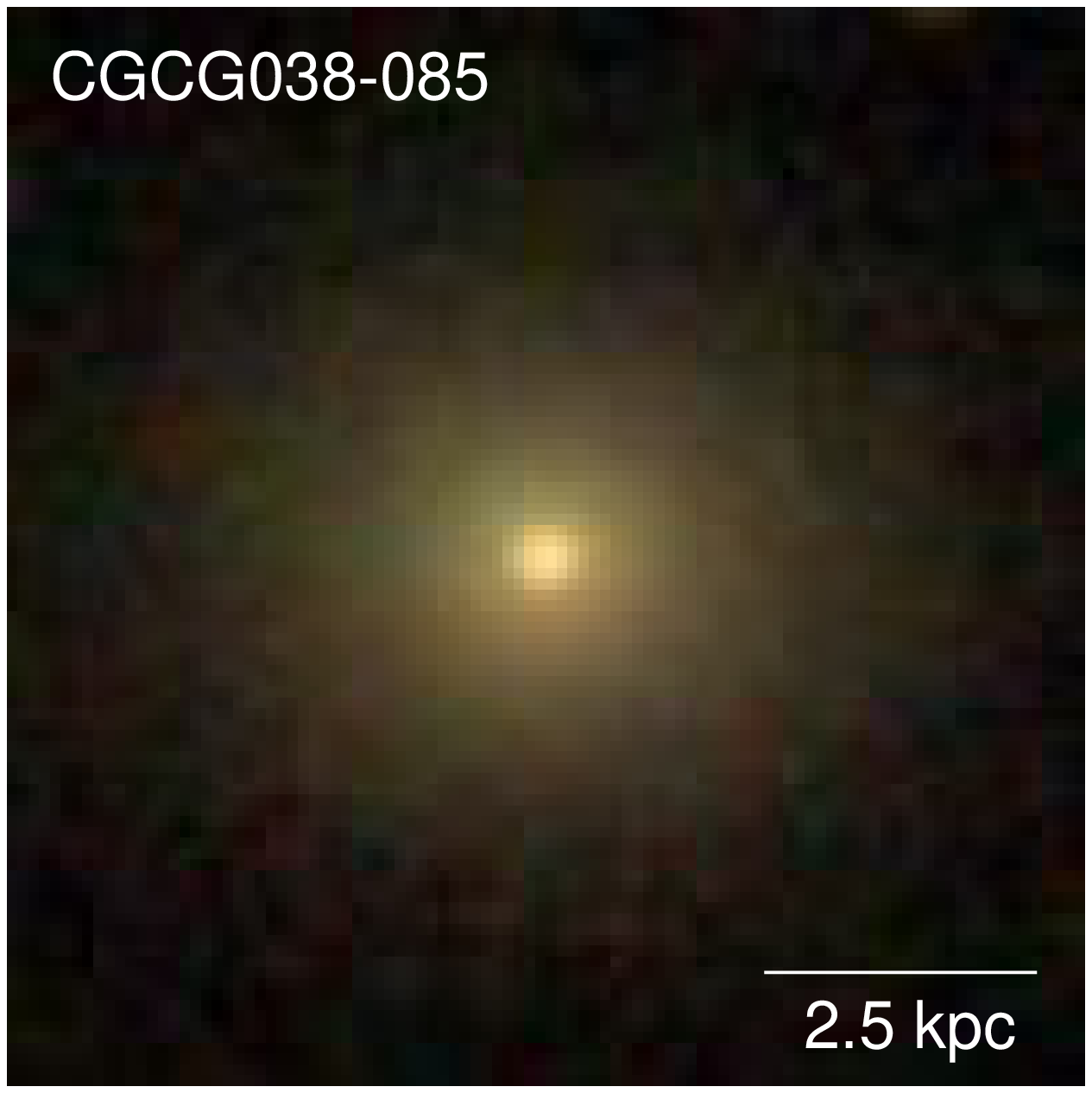} \\
   \includegraphics[width=0.33\textwidth,angle=270]{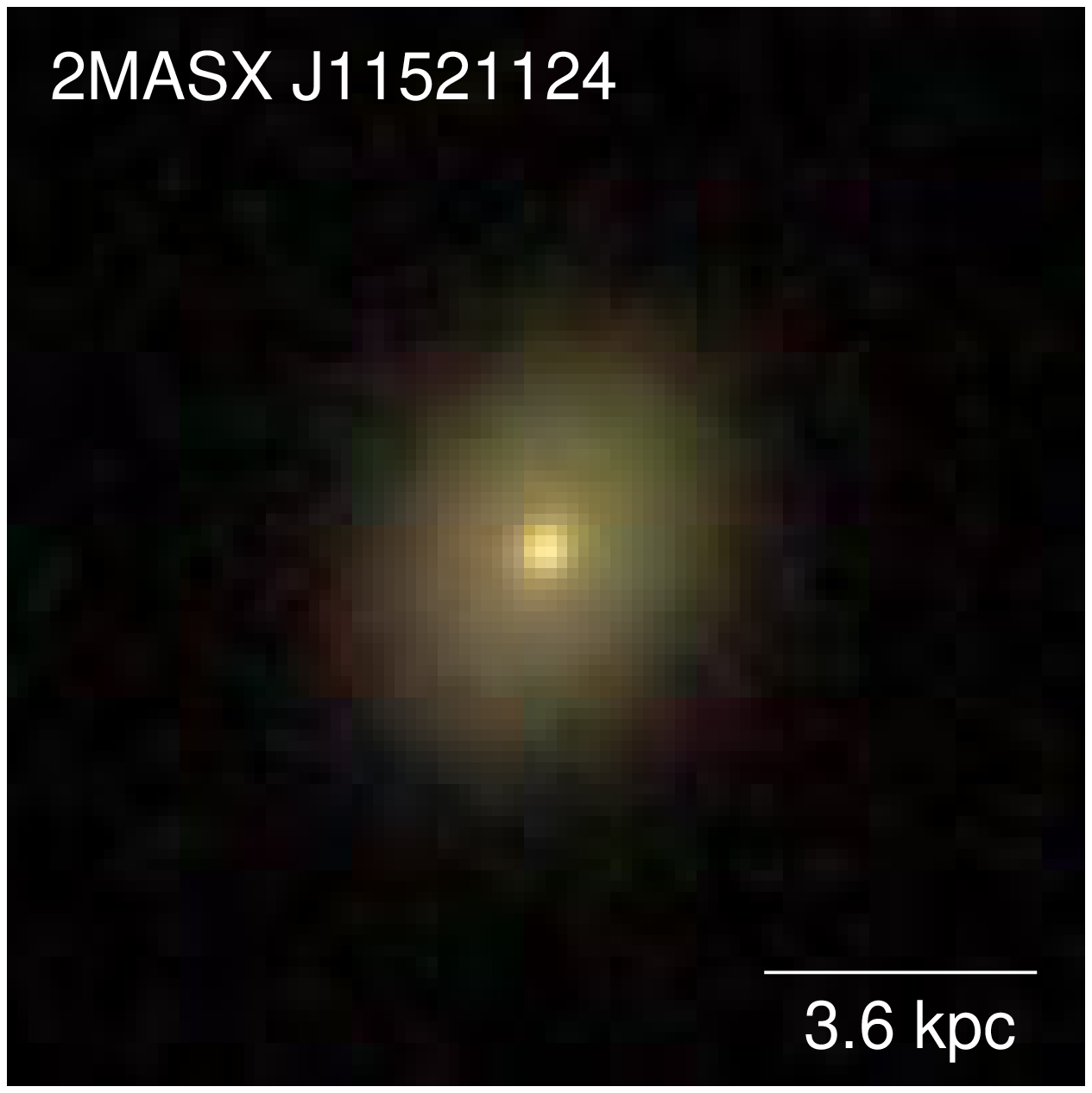} 
   \includegraphics[width=0.33\textwidth,angle=270]{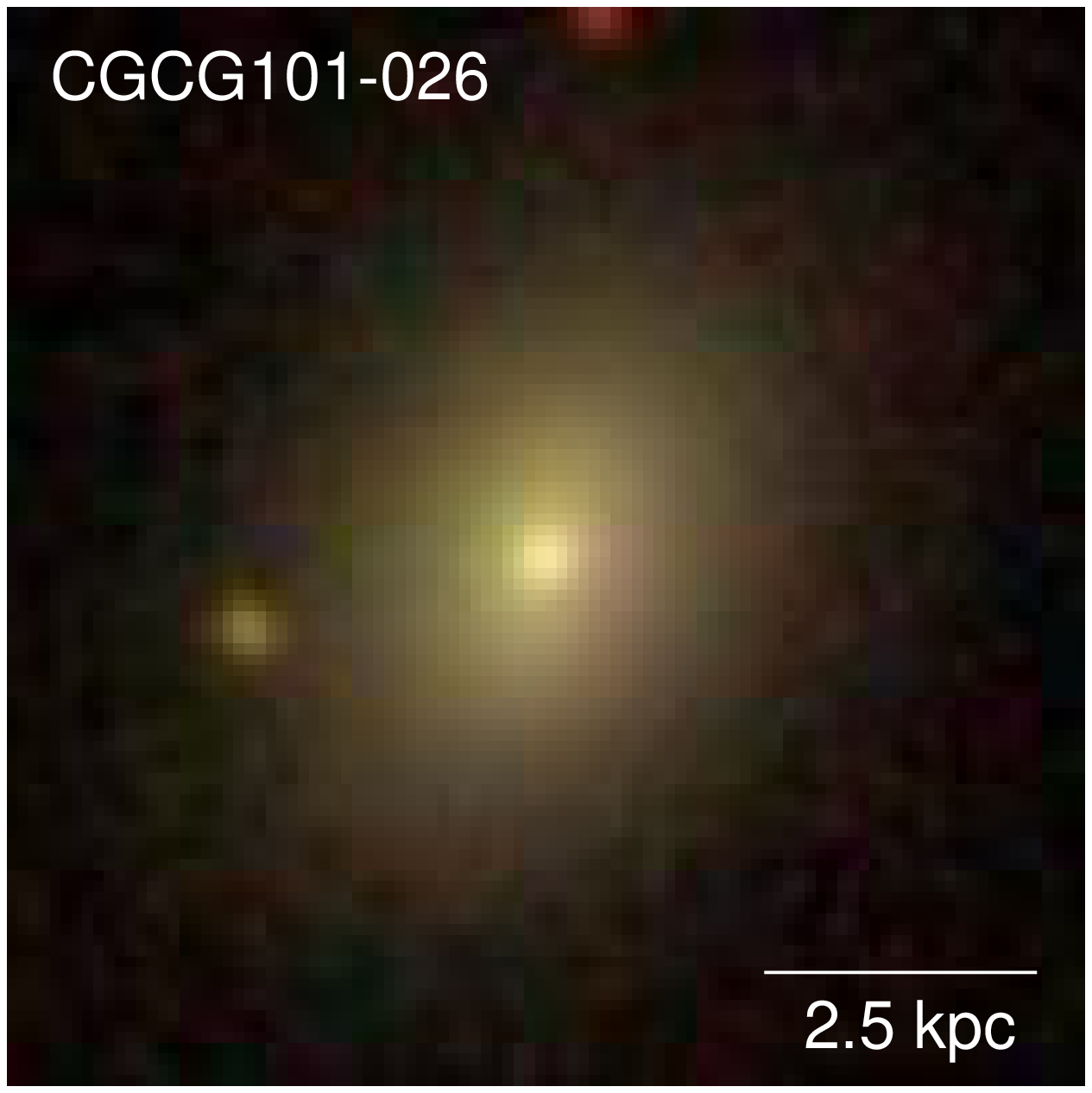} 
   \includegraphics[width=0.33\textwidth,angle=270]{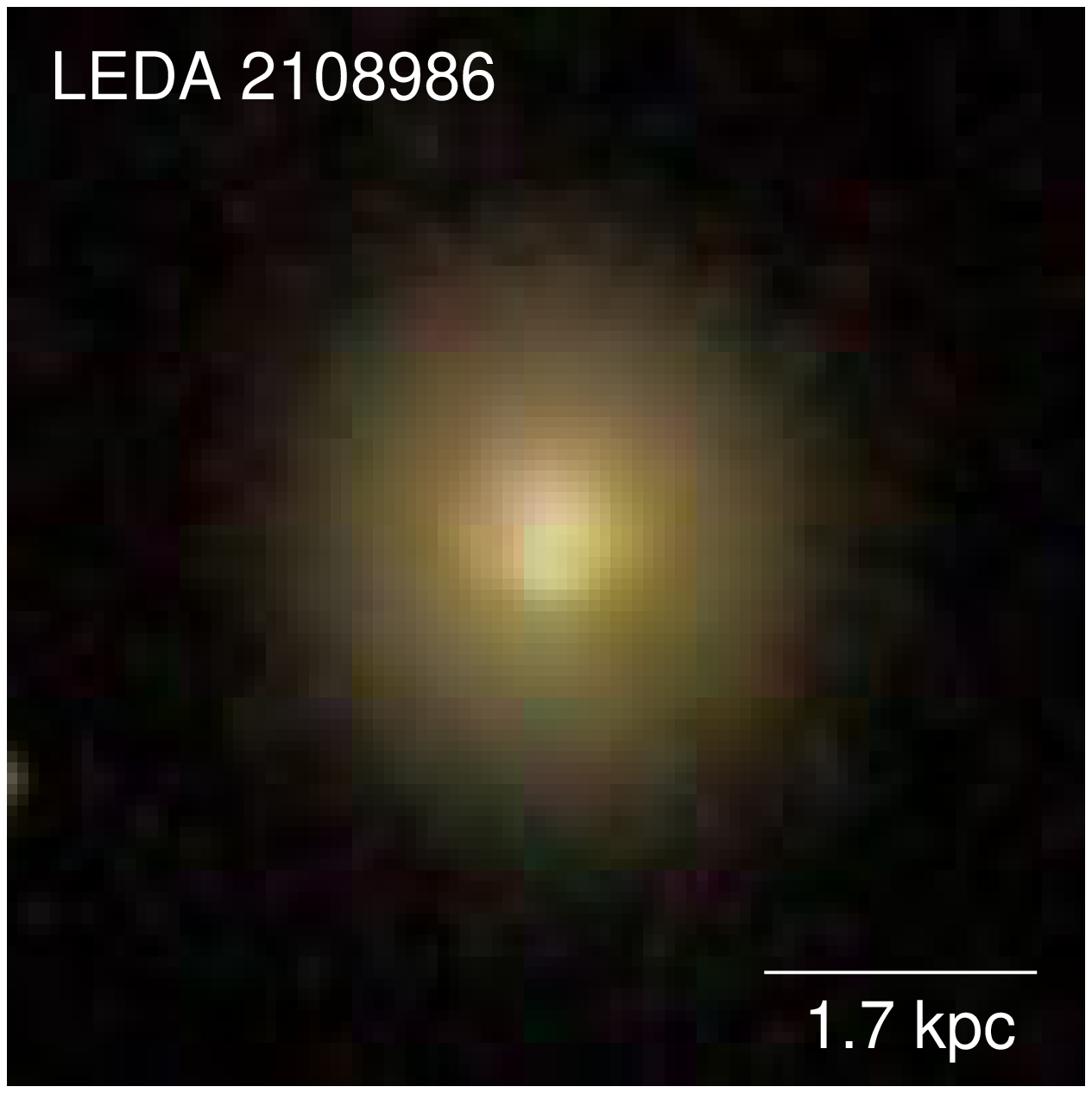}       
\caption{SDSS $r$-band images of the galaxies in our spectroscopic sample. An angular scale of 10\arcsec\  is provided in the lower right corner of each panel together with the corresponding physical scale. North is up and East is left. }
   \label{fig:images}
\end{figure*}

{ From this sample of quenched low-mass galaxies we select those galaxies that are well separated ($D_\textrm{proj}\gtrsim1$~Mpc) from massive neighbours  (Table~\ref{table:specdEs}). }
Both the SDSS and the 2MASS redshift survey \citep[][]{2012ApJS..199...26H} were queried for bright neighbours 
 with $M_{K_s} < -23$~mag { (see \citealt{2012ApJ...757...85G}; corresponding to $M_{\star}\gtrsim 3 \times 10^{10}$~M$_{\odot}$ when assuming a $K_s$-band mass-to-light ratio of unity)}
within  a velocity interval of $\pm500$~km~s$^{-1}$ around the recession velocity of each low-mass galaxy.
The projected distance between the quenched dwarfs and their nearest bright neighbour was  converted to a physical scale assuming
the distance of the dwarf given by its recession velocity.
{ These restrictions are slightly less restrictive than those criteria for isolation used by \citet{2012ApJ...757...85G}.
We discuss the differences, and also a more restrictive choice of the massive neighbours luminosity} ($M_{K_s} < -21.5${~mag) in Section \ref{sec:iso}.}
 
 { The closest neighbour distance limit excluded the majority of quenched low-mass galaxies (mean projected separation to nearest bright neighbour $D_{\rm proj}=0.36$ Mpc; Fig.~\ref{fig:dhist}), which are located in groups and clusters, and resulted in a sample of 46 candidates of isolated, quenched dwarf galaxies (highlighted as blue points in Fig.~\ref{fig:psample}). 
 { The median distance in projection to the closest bright neighbour beyond 1 Mpc for this sample is 1.85 Mpc, and the median velocity difference relating  to that neighbour is $| \Delta V | =222$ km~s$^{-1}$. }

From this sample suitable targets were chosen so that their surface brightness allowed us to measure extended rotation curves with reasonable integration times, and that the observations were not hampered by bright foreground or background objects. 
For comparison we also calculate their stellar masses from the absolute
$r$-band magnitudes, assuming a mean stellar age of 5 Gyr (comparable to
quenched dwarfs in the Virgo Cluster, see \citealt{2014ApJS..215...17T}),  $[\textrm{Fe}/\textrm{H}] = -0.33$ and a Kroupa initial mass function (see Table~\ref{table:specdEs}).}
The median stellar mass for the spectroscopic sample is $M_\star \sim 4\times10^9\, \textrm{M}_\odot$.
Fig.~\ref{fig:images} shows the nine galaxies   (upper part of Table~\ref{table:specdEs}) for which we obtained spectra.

We note that one of our targets (CGCG038-085) was classified as a member of a galaxy group by \citet{2006ApJS..167....1B} and has a neighbouring galaxy with a angular separation of $48\arcsec$ corresponding to 36 kpc at the group distance.  The recession velocity of the target listed by NED suggests it is a group member. However, the SDSS spectrum and our analysis place the object clearly in the foreground with respect to the group ($\Delta V > 7000$\,km\,s$^{-1}$ with a velocity dispersion of the group of 57\,km\,s$^{-1}$).
\citet{2012AJ....144...57F} list VIIIZw040 and LEDA 2108986 as extremely isolated early-type galaxies. The study mentions  two companion galaxies fainter than $M_V = -16.5$~mag for the latter.
While we could not determine with certainty, which galaxies were meant, we found 
the two nearest neighbours (with $M_r=-15.28$~mag and $M_r = -15.78$~mag) to be well separated with projected distances of 2.0 and 2.1 Mpc, respectively. Other galaxies with smaller angular separations are in background.
CGCG101-026, LEDA 2108986, and VIIIZw040 are also listed as isolated galaxies by \citet{2015A&A...578A.110A}, with separations from their closest neighbour of 2.01, 2.03, and 2.41 Mpc, respectively. In these cases, the closest neighbouring galaxies are fainter than our limiting magnitude of $M_{K_s} < -23$~mag.

\begin{table*}
\begin{center}
\caption{Sample of isolated quenched low-mass galaxies. \label{table:specdEs}}
\scalebox{0.85}{
\begin{tabular}{lcccccccccc}
\hline \hline
Galaxy         & PGC &  RA  &  Dec &       Type  & $M_{B,\textrm{tot}}$ & $\log_{10} (M_\star/\textrm{M}_\odot)$&   \multicolumn{4}{c}{closest luminous neighbour}\\
& & & &  & & & $D_{500}$ & $|\Delta V_{500}|$ & $D_{1000}$ & $|\Delta V_{1000}|$  \\
 	       &    & [$h$:$m$:$s$]  & [$d$:$m$:$s$]  &   & [mag] & & [Mpc] &  [km s$^{-1}$]  &  [Mpc] & [km s$^{-1}$]  \\
	       \hline
LEDA~3115955 & 3115955 & 02:05:46.37 & +00:37:33.8  & E & $-18.10\pm 0.35$ & 9.73 &  1.67 &   13 \\
2MASX~J03190758+4232179 & 2200915 & 03:19:07.58 & +42:32:17.9  & -- & $-17.41\pm 0.34$ & 9.43 &  0.94 &  130 \\
LCSBS1123P & 1719550 & 08:17:15.93 & +24:53:56.8  & E & $-15.17\pm 0.27$ & 8.61 &  1.17 &  152 \\
2MASX~J08192430+2100125 & 23346 & 08:19:24.30 & +21:00:12.5  & E? & $-17.51\pm 0.41$ & 9.75 &  1.00 &  314  &  0.11 &  987 \\
VIIIZw040 & 25889 & 09:11:05.58 & +09:20:58.5  & E(R) & $-17.63\pm 0.37$ & 9.52 &  3.78 &  455 \\
CGCG038-085 & 33187 & 11:00:41.76 & +04:07:29.6  & E & $-17.73\pm 0.40$ & 9.63 &  2.03 &   75 \\
2MASX~J11521124+0421239 & 1267103 & 11:52:11.24 & +04:21:23.9  & E & $-17.62\pm 0.43$ & 9.58 &  4.48 &  292  &  1.38 &  662 \\
CGCG101-026 & 46516 & 13:20:06.92 & +14:32:35.7  & ES0 & $-18.04\pm 0.25$ & 9.68 &  2.40 &  263 \\
LEDA~2108986 & 2108986 & 15:03:15.57 & +37:45:57.2  & ES0 & $-16.96\pm 0.38$ & 9.24 &  3.07 &   21 \\
\hline
J001212.41--110010.4 & 3099480 & 00:12:12.41 & --11:00:10.4  & -- & $-16.04\pm 0.36$ & 8.77 &  3.71 &  387  &  1.51 &  509 \\
J001530.03+160429.7 & 1500123 & 00:15:30.03 & +16:04:29.7  & -- & $-15.91\pm 0.40$ & 8.91 &  1.63 &  211 \\
J001601.19+160133.4 & 212493 & 00:16:01.19 & +16:01:33.4  & Sb & $-17.67\pm 0.45$ & 9.62 &  1.76 &   41 \\
J012506.69--000807.0 & 4131836 & 01:25:06.69 & --00:08:07.0  & -- & $-15.58\pm 0.39$ & 8.67 &  1.35 &   31 \\
J013842.89--002053.0 & 1145615 & 01:38:42.89 & --00:20:53.0  & E & $-16.75\pm 0.40$ & 9.15 &  1.54 &   87 \\
J045058.77+261313.7 & 1767757 & 04:50:58.77 & +26:13:13.7  & -- & $-15.82\pm 0.40$ & 9.84 &  1.94 &  280  &  1.83 &  661 \\
J075303.96+524435.8 & 3724966 & 07:53:03.96 & +52:44:35.8  & E & $-17.21\pm 0.50$ & 9.38 &  1.04 &   63 \\
J082013.92+302503.0 & 1902931 & 08:20:13.92 & +30:25:03.0  & E & $-17.23\pm 0.35$ & 9.37 &  2.94 &  102 \\
J082210.66+210507.5 & 1642425 & 08:22:10.66 & +21:05:07.5  & -- & $-15.40\pm 0.41$ & 8.72 &  1.02 &  453  &  0.36 &  551 \\
J084915.01+191127.3 & 4572078 & 08:49:15.01 & +19:11:27.3  & -- & $-15.67\pm 0.29$ & 8.59 &  3.72 &  373  &  0.11 &  526 \\
J085652.63+475923.8 & 2308331 & 08:56:52.63 & +47:59:23.8  & E & $-17.29\pm 0.34$ & 9.34 &  3.67 &  277 \\
J091514.45+581200.3 & 26096 & 09:15:14.45 & +58:12:00.3  & E & $-17.66\pm 0.38$ & 9.57 &  6.90 &  350  &  3.61 &  714 \\
J091657.98+064254.2 & 3456113 & 09:16:57.98 & +06:42:54.2  & E & $-17.55\pm 0.35$ & 9.52 &  2.56 &  211 \\
J093016.38+233727.9 & 1690666 & 09:30:16.38 & +23:37:27.9  & E & $-16.83\pm 0.28$ & 9.19 &  2.27 &  416 \\
J093251.11+314145.0 & 3743536 & 09:32:51.11 & +31:41:45.0  & E & $-17.74\pm 0.35$ & 9.52 &  3.30 &  191 \\
J094408.52+111514.8 & 3531658 & 09:44:08.52 & +11:15:14.8  & Sb & $-17.91\pm 0.35$ & 9.63 &  2.77 &  320  &  0.34 &  922 \\
J094834.50+145356.6 & 1469954 & 09:48:34.50 & +14:53:56.6  & E & $-17.80\pm 0.59$ & 9.84 &  1.41 &    6 \\
J100003.93+044845.0 & 1273634 & 10:00:03.93 & +04:48:45.0  & E & $-15.76\pm 0.37$ & 8.88 &  1.24 &   80 \\
J105005.53+655015.6 & 3097669 & 10:50:05.53 & +65:50:15.6  & E & $-16.68\pm 0.30$ & 9.01 &  5.66 &  400 \\
J110423.34+195501.5 & 3765330 & 11:04:23.34 & +19:55:01.5  & E & $-15.43\pm 0.35$ & 8.62 &  1.07 &   11 \\
J112422.89+385833.3 & 4098131 & 11:24:22.89 & +38:58:33.3  & -- & $-12.94\pm 0.50$ & 7.87 &  1.30 &  354  &  0.06 &  925 \\
J114423.13+163304.5 & 213877 & 11:44:23.13 & +16:33:04.5  & E & $-17.88\pm 0.37$ & 9.71 &  1.47 &  386  &  0.02 &  549 \\
J120300.94+025011.0 & 1238733 & 12:03:00.94 & +02:50:11.0  & E & $-17.46\pm 0.52$ & 9.72 &  1.06 &   55 \\
J120823.99+435212.4 & 2231154 & 12:08:23.99 & +43:52:12.4  & E & $-17.61\pm 0.32$ & 9.66 &  4.42 &  230 \\
J122543.23+042505.4 & 1267954 & 12:25:43.23 & +04:25:05.4  & Sb & $-17.80\pm 0.38$ & 9.66 &  3.24 &  261  &  2.08 &  804 \\
J124408.62+252458.2 & 1735700 & 12:44:08.62 & +25:24:58.2  & E & $-17.61\pm 0.49$ & 9.62 &  1.11 &  421 \\
J125026.61+264407.1 & 1786738 & 12:50:26.61 & +26:44:07.1  & E & $-16.72\pm 0.31$ & 9.04 &  1.05 &  353 \\
J125103.34+262644.6 & 94058 & 12:51:03.34 & +26:26:44.6  & E & $-17.61\pm 0.30$ & 9.46 &  1.21 &  180 \\
J125321.68+262141.1 & 1773332 & 12:53:21.68 & +26:21:41.1  & Sc & $-17.54\pm 0.38$ & 9.40 &  1.05 &  382 \\
J125756.52+272256.2 & 126848 & 12:57:56.52 & +27:22:56.2  & E & $-16.58\pm 0.29$ & 9.12 &  1.17 &  215  &  0.32 &  836 \\
J125940.10+275117.7 & 44654 & 12:59:40.10 & +27:51:17.7  & S0a & $-16.94\pm 0.29$ & 9.45 &  4.20 &  123  &  0.23 &  790 \\
J130320.35+175909.7 & 4352778 & 13:03:20.35 & +17:59:09.7  & E & $-13.05\pm 0.02$ & 7.73 &  1.04 &  164  &  0.96 &  553 \\
J130549.09+262551.6 & 1775990 & 13:05:49.09 & +26:25:51.6  & E & $-16.95\pm 0.31$ & 9.22 &  2.02 &   39 \\
J142914.46+444156.3 & 51753 & 14:29:14.46 & +44:41:56.3  & S0 & $-17.81\pm 0.28$ & 9.60 &  2.51 &  243 \\
J144621.10+342214.1 & 52741 & 14:46:21.10 & +34:22:14.1  & S0 & $-17.58\pm 0.32$ & 9.63 &  6.25 &  460  &  3.89 &  926 \\
J160810.70+313055.0 & 1950976 & 16:08:10.70 & +31:30:55.0  & E & $-17.74\pm 0.41$ & 9.55 &  5.52 &   72 \\
J232028.21+150420.8 & 1474455 & 23:20:28.21 & +15:04:20.8  & -- & $-17.61\pm 0.34$ & 9.53 &  1.02 &  438  &  0.84 &  559 \\
\hline
 \end{tabular}}
\end{center}
\parbox{0.95\textwidth}{{\bf Notes:   }  
The first four columns list the galaxy name (for the objects with the 2MASX designation we use abridged versions of the name throughout the paper), its number in the PGC, and its coordinates (equinox J2000). The galaxy classifications (column 5) and total $B$-band magnitudes (column 6; extinction corrected) are from HyperLEDA. The distances for the absolute magnitudes are calculated using the recession velocities (from SDSS and our updated values for the galaxies above the line).
{  Column 7 lists the stellar mass (see text).}
The closest bright neighbour within a velocity interval of $\pm 500$ km~s$^{-1}$ is extracted from a combined catalogue containing redshifts from the NASA Sloan Atlas and the 2MASS redshift survey.  
The projected (linear) distance between the neighbour and the low-mass galaxy and the difference between their recession velocities are given in columns 8 and 9.
{ If there is an additional closer neighbour within a velocity interval of $\pm 1000$ km~s$^{-1}$, the corresponding details are listed in columns 10 and 11.}
Keck ESI spectroscopy is presented for the galaxies above the line in Section~\ref{sec:kins}. }
\end{table*}

\section{Data}
\label{section:data}

\begin{table}
\begin{center}
\caption{Observing log. \label{table:obslog}}
\begin{tabular}{lcccc}
\hline \hline
Galaxy              &  Run & Exp. time & PA & Seeing \\
 			& &       [min]  & [deg] & [arcsec]  \\
\hline
LEDA 3115955    & 15B & 90  & 39  & 1.0 \\ 
2MASX J03190758 & 15B & 60  & 72  & 0.9\\
LCSBS1123P     & 15B & 120 & 178 & 1.1 \\
2MASX J08192430 & 15B & 90  & 50  & 0.8\\ 
VIIIZw040      & 16A & 40  & 14  & 0.8 \\
CGCG038085     & 16A & 120 & 84  & 0.7\\
2MASX J11521124 & 15B & 90  & 156 & 0.8\\
CGCG101026     & 16A & 60  & 159 & 0.9\\
LEDA 2108986    & 16A & 75  & 158 & 0.8 \\
\hline
\end{tabular}
\parbox{0.45\textwidth}{{\bf Notes:   }   The second column indicates during which run the spectra were taken, the third  and fourth column list the total integration time and the 
position angle of the slit (PA; East of North), while the fifth column gives the FWHM of the seeing. During the 16A run the conditions were variable.}
\end{center}
\end{table}

We observed 9 of  the quenched, isolated low-mass galaxies (Table~\ref{table:specdEs}) with the  Echellette Spectrograph and Imager \citep[ESI,][]{Sheinis:2002ft} on the Keck II telescope  
on 2016 January 11th and 2016 March 14th.
ESI was used in the echellette mode with a slit width of $0\farcs75$. This setup results in a spectral coverage of $\sim$4000 to 10000\,\AA{} across ten echelle orders,
and a velocity resolution of $\sigma\sim$25\,km\,s$^{-1}$.
The galaxy centre was put in the middle of the slit which was aligned with the galaxy's major axis.
The integrations for each object were split into at least three individual exposures. 
{ Total exposure times, position angles, and observing conditions  are summarised in Table \ref{table:obslog}.}
Furthermore, we obtained several stellar spectra with the same setup. 
 In each run we observed one star to trace the echelle orders in the reduction process,
as well as velocity standards to be used as templates for measuring the kinematics. This template library was augmented by spectra of velocity standards
from previous runs so that our stellar library comprises in total 16  standards, spanning spectral types from O9V to K7V  with a range of metallicities and containing larger numbers of 
standards for the later spectral types.

We prepared master frames of the necessary calibrations (internal flat fields, bias, arc lamps) 
by combining the two-dimensional spectra  with the {\tt imcombine} task in IRAF.
The subsequent data reduction was then carried out  with  {\sc makee}.\footnote{Written by T.~Barlow, \url{http://www2.keck.hawaii.edu/inst/esi/makee.html}.}
The pipeline subtracts the bias from the science spectra and divides them by the flat field.
It traces the echelle orders with a bright trace star and extracts one-dimensional spectra for each order
in spatial apertures that can be defined by the user.

To obtain spatially resolved kinematics, we extracted spectra from apertures  with a width of $0\farcs9$ along the slit,
which corresponds to the average FWHM of the seeing.
The first extraction is centred on the galaxy, while the next ones were extracted from increasingly  larger radii to both
sides. The largest angular extent was probed for 2MASX~J08192430, for which we extracted 5 spectra on each side of the
central spectrum. In this case the apertures furthest away from the galaxy centre  are twice as wide as the
other to improve the signal-to-noise ($S/N$). 

\begin{figure*}  
   \centering
   \includegraphics[height=0.33\textwidth,angle=-90]{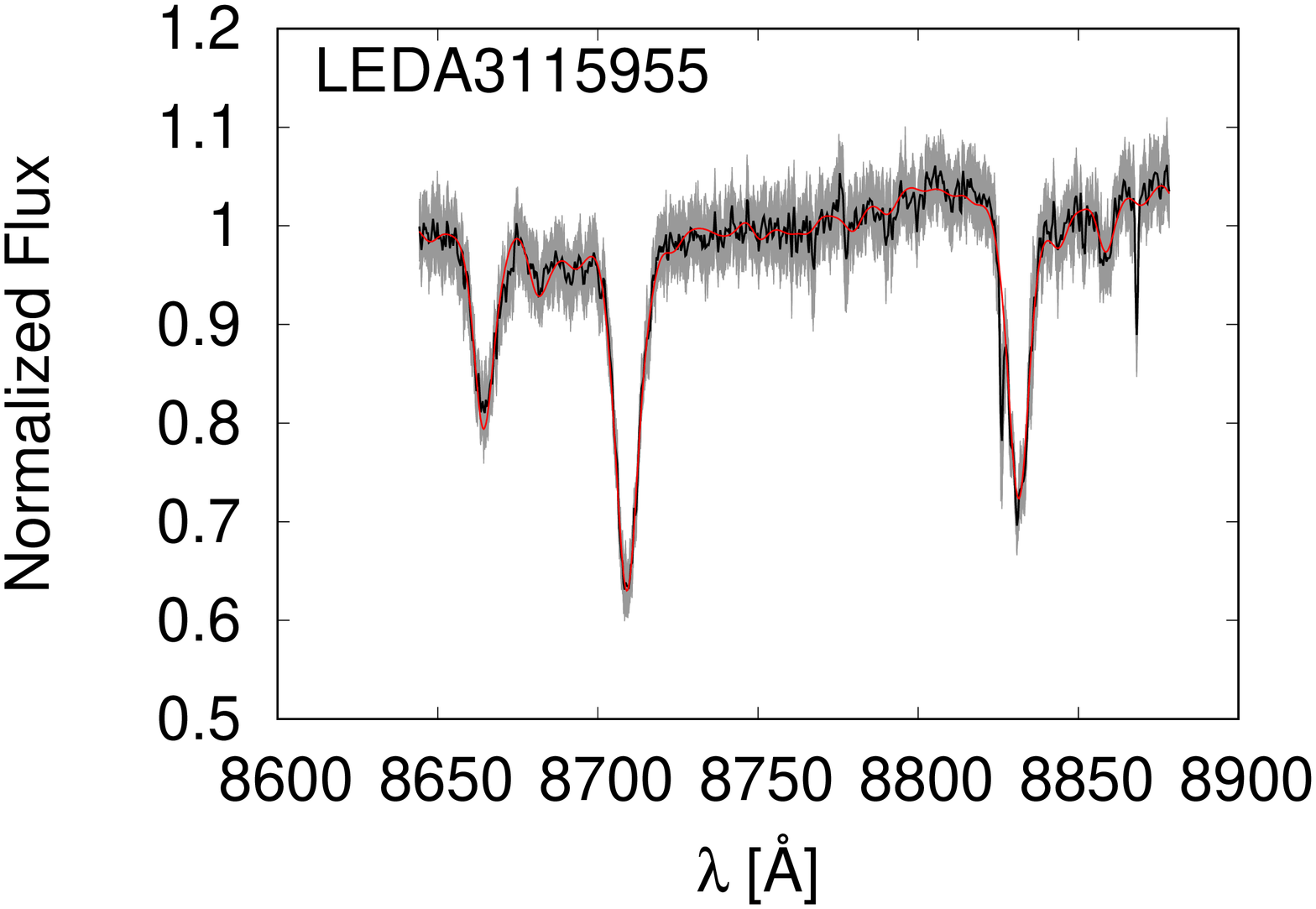} 
   \includegraphics[height=0.33\textwidth,angle=-90]{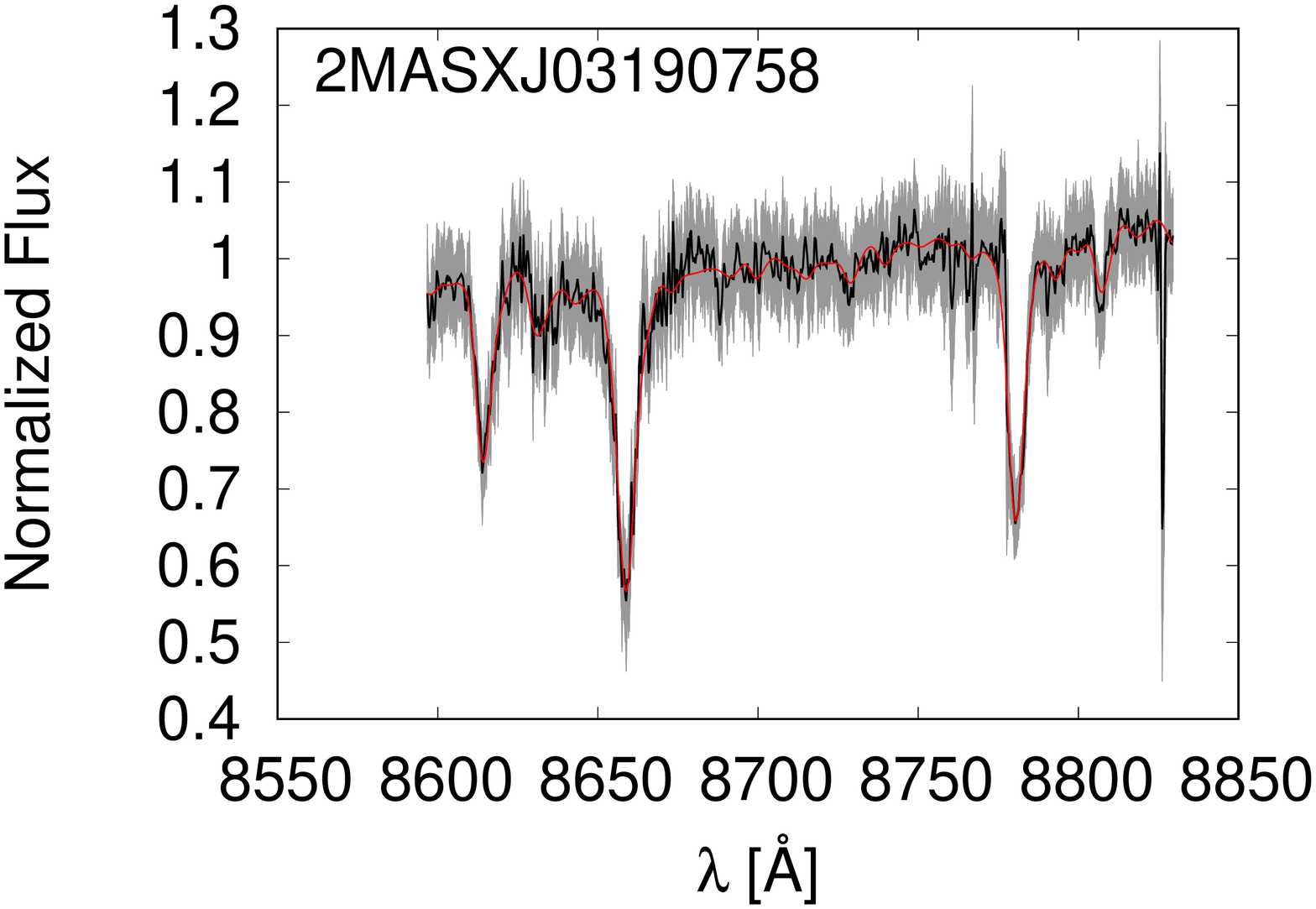} 
   \includegraphics[height=0.33\textwidth,angle=-90]{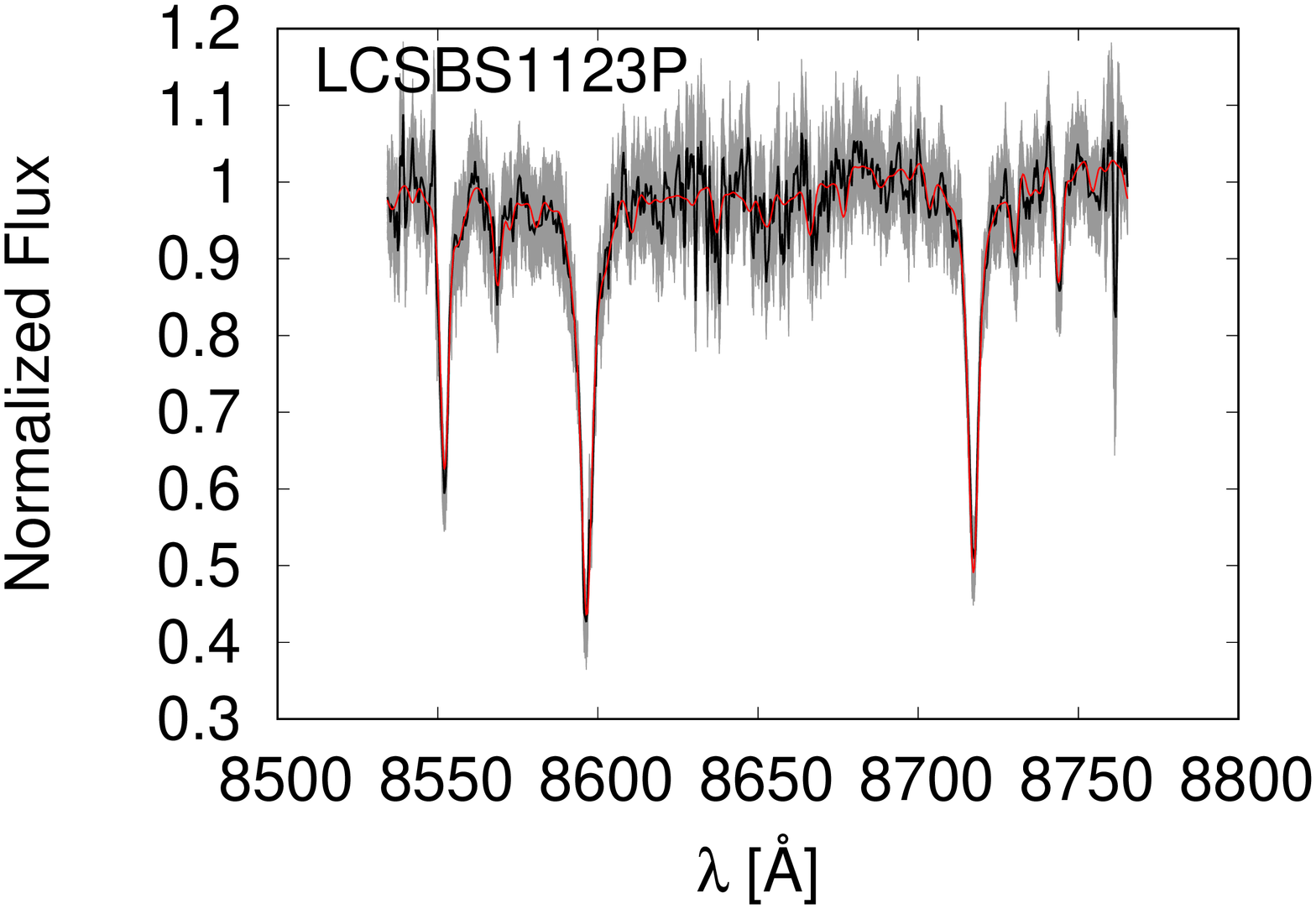} \\
   \includegraphics[height=0.33\textwidth,angle=-90]{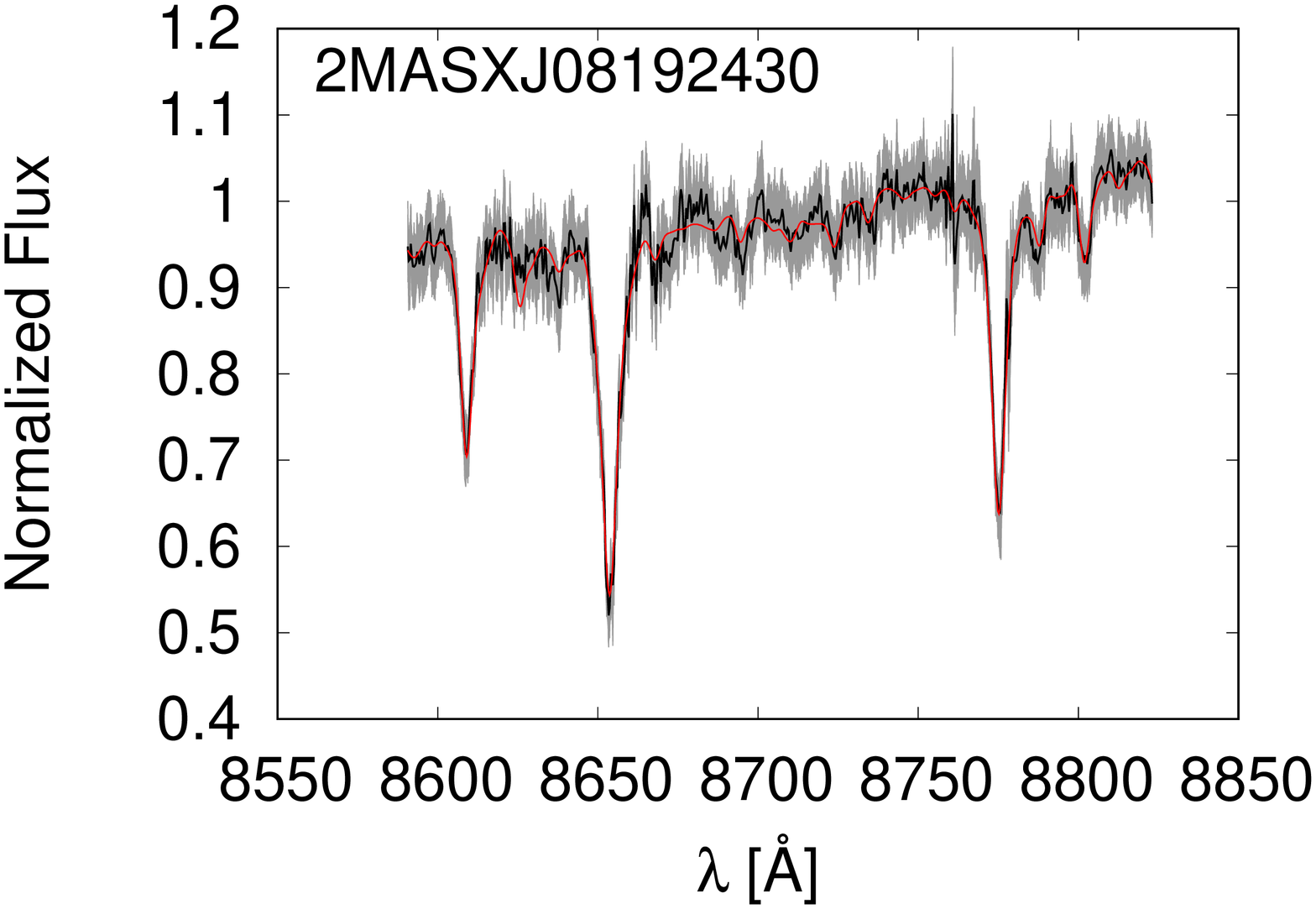} 
   \includegraphics[height=0.33\textwidth,angle=-90]{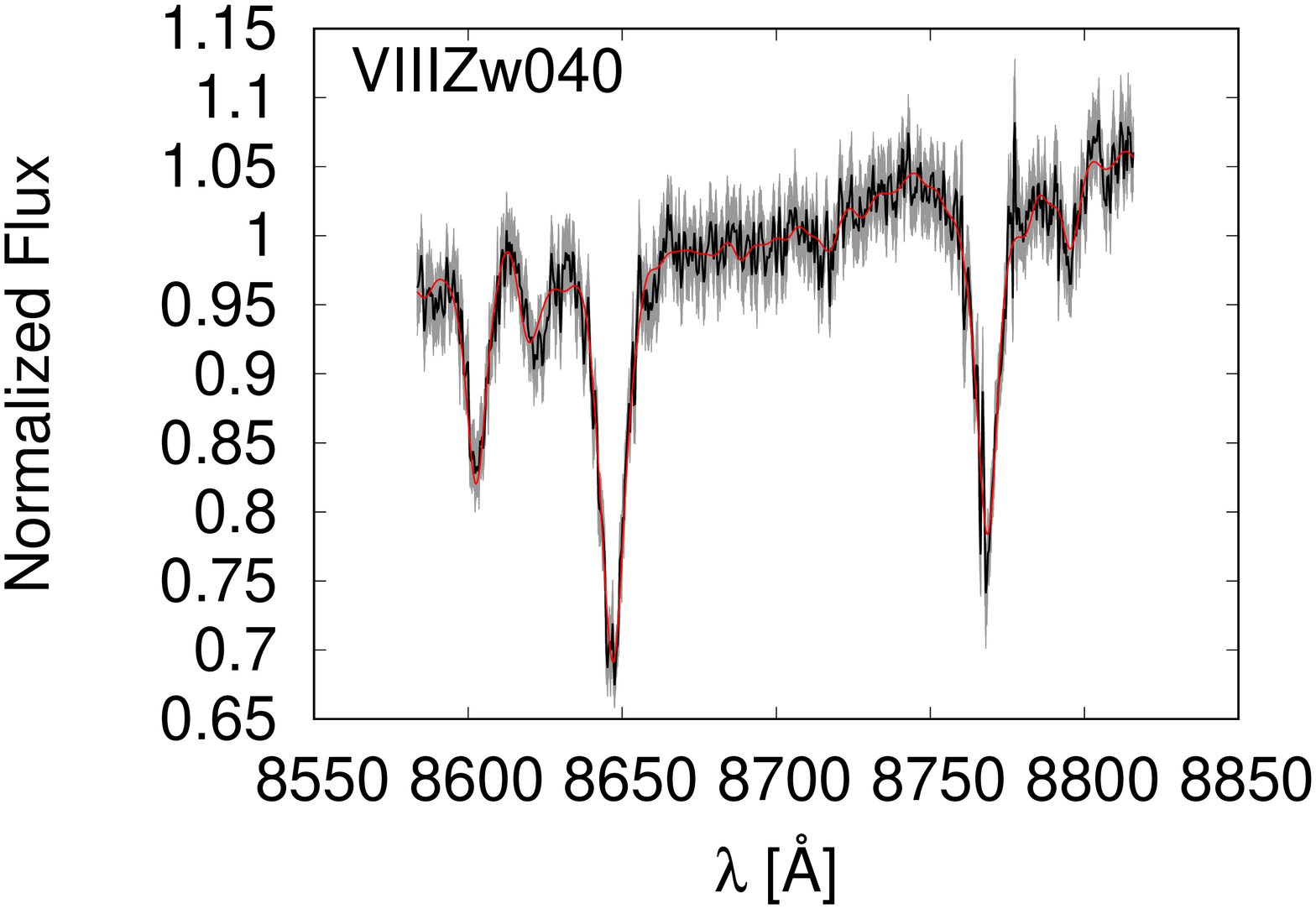} 
   \includegraphics[height=0.33\textwidth,angle=-90]{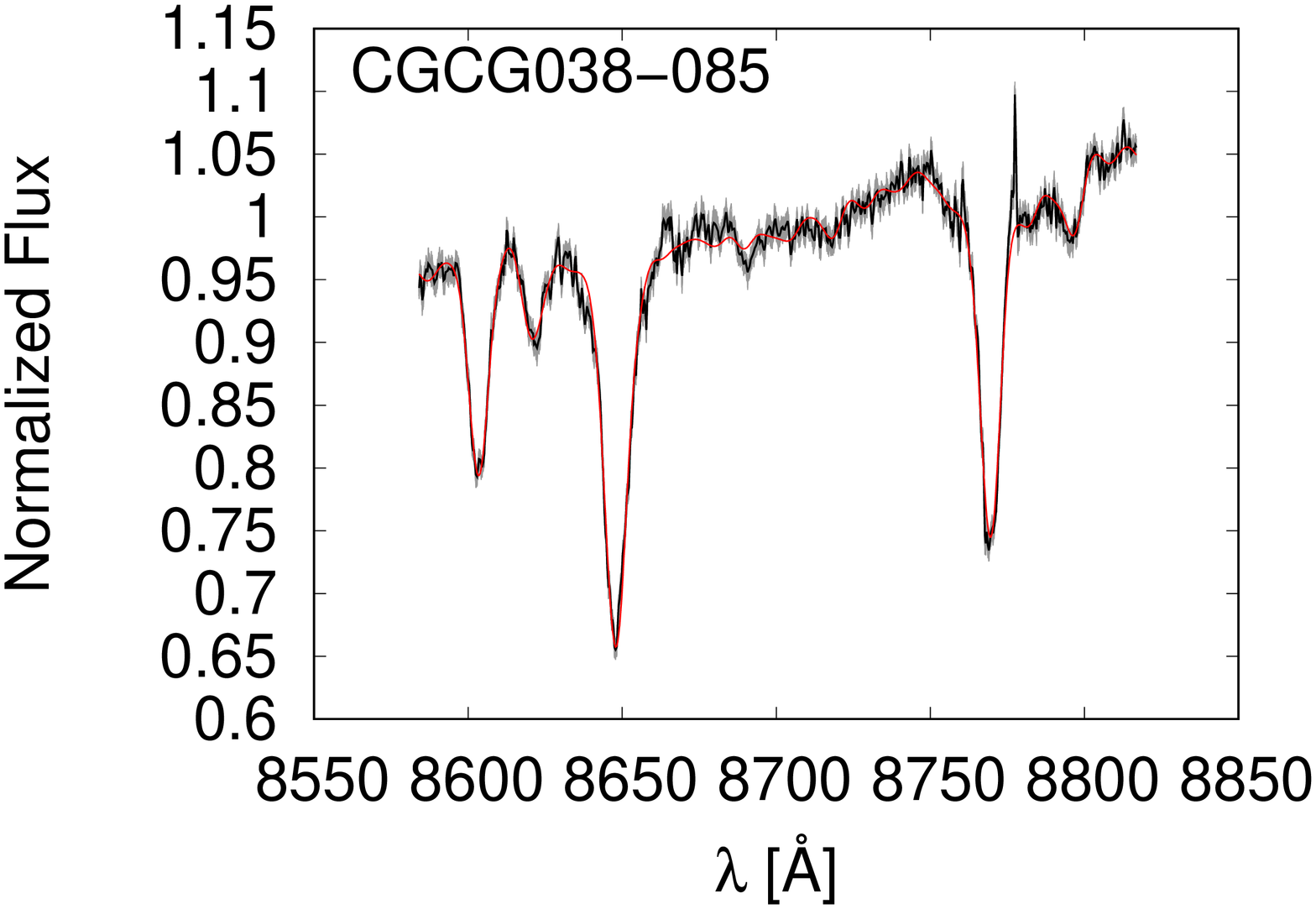} \\
   \includegraphics[height=0.33\textwidth,angle=-90]{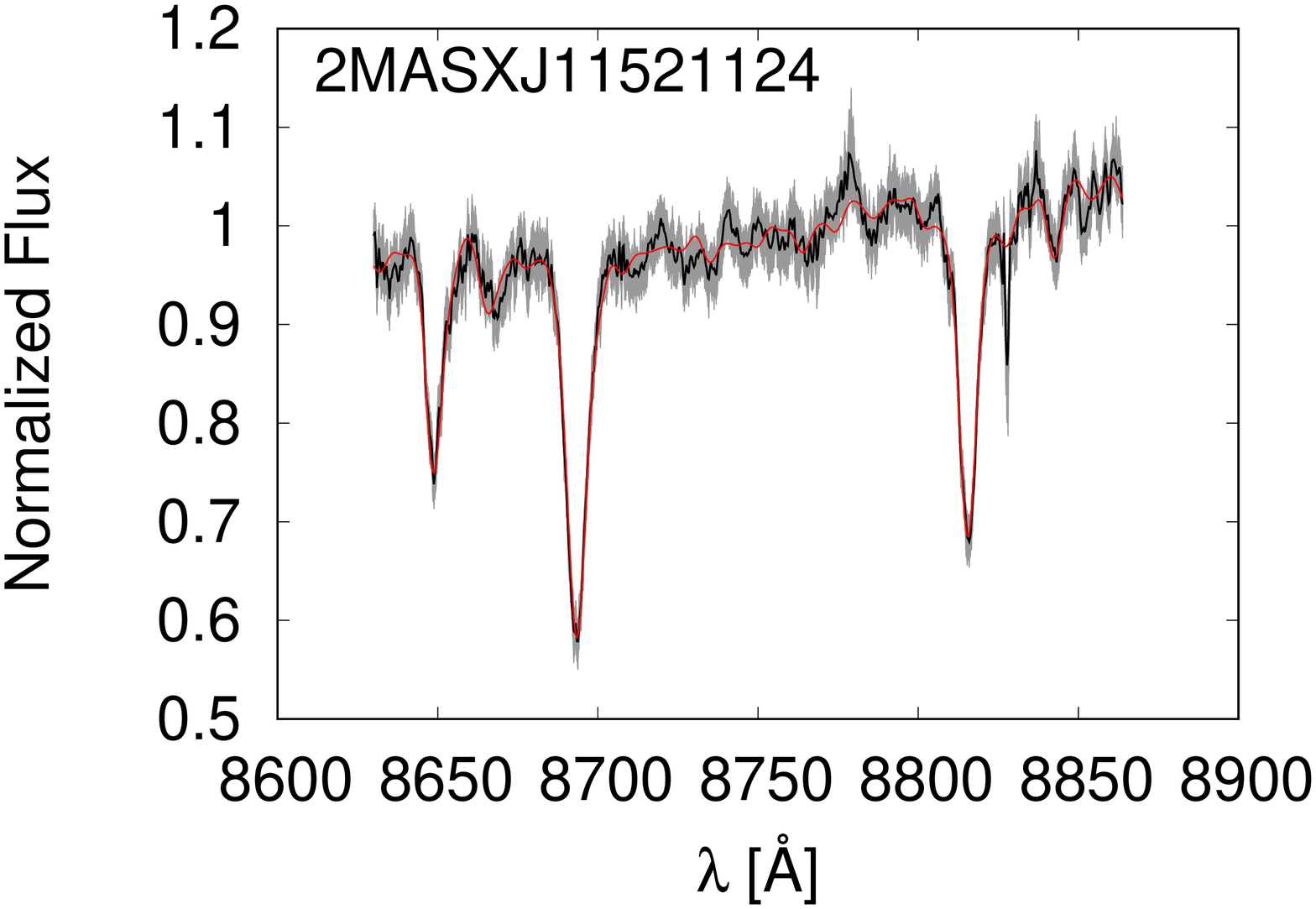} 
   \includegraphics[height=0.33\textwidth,angle=-90]{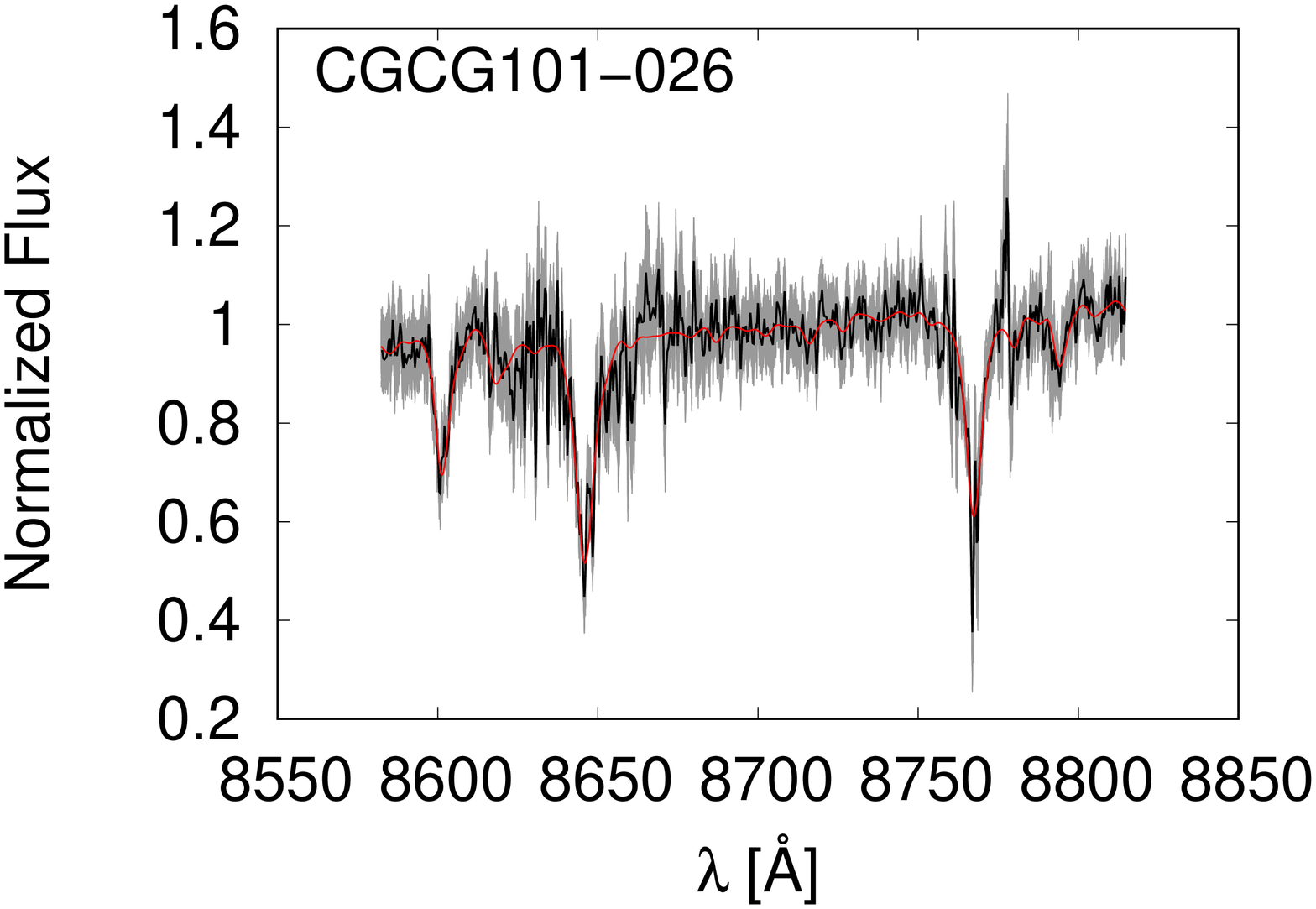} 
   \includegraphics[height=0.33\textwidth,angle=-90]{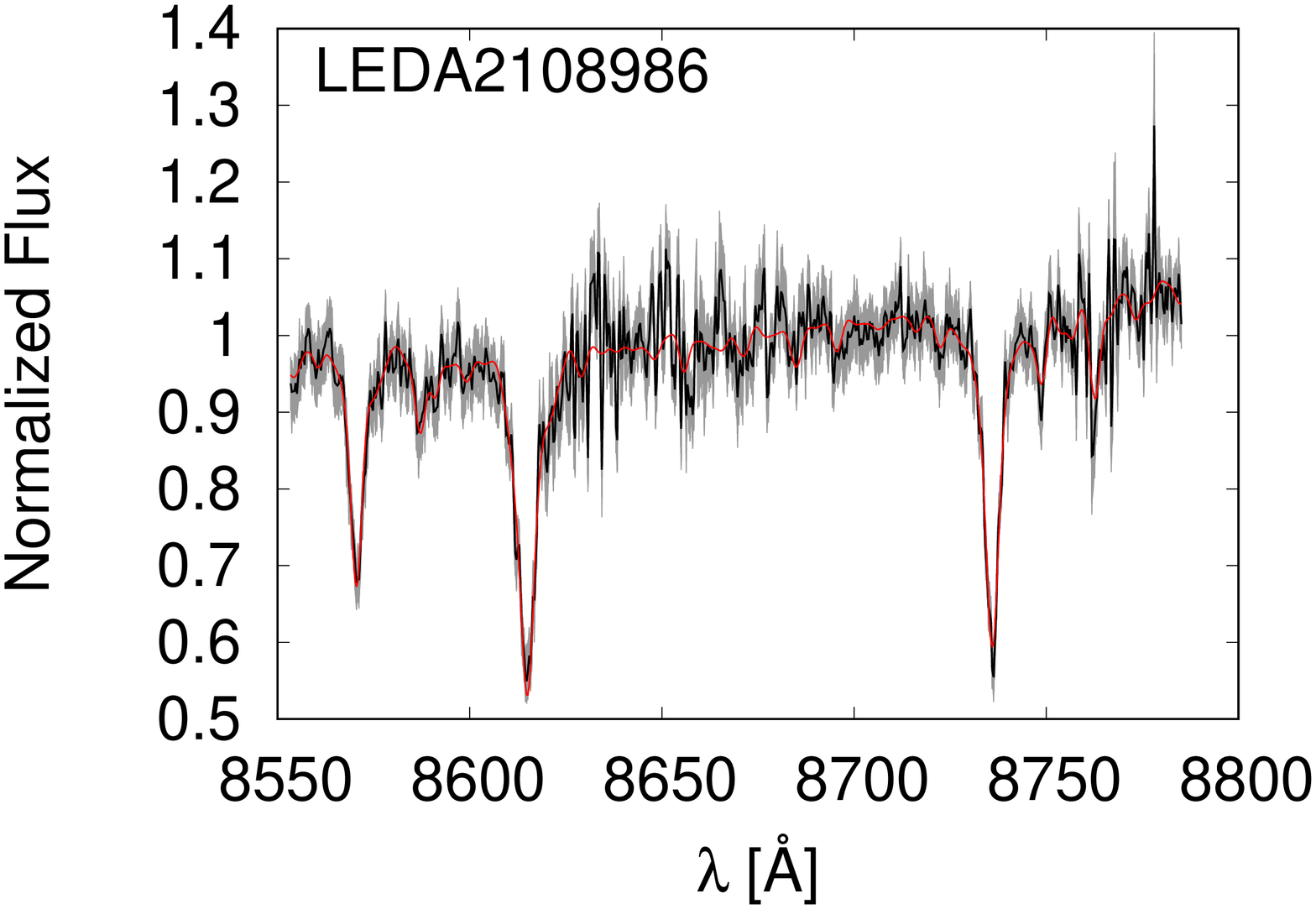} 
      \caption{Spectra of the central aperture for all galaxies in the Ca\,{\sc ii} triplet spectral region. The black and red lines are the observed spectrum and the \textsc{pPXF} fit, respectively,
      while the Monte Carlo realisations of the observed spectra are plotted in grey.}
   \label{fig:spec}
\end{figure*}

The spectra were wavelength calibrated using the known emission lines of the arc lamps and the night sky.
The pixel uncertainties  were calculated for each of the spectra. Finally, the extracted spectra from individual exposures
were coadded for each aperture in a $S/N$ optimised manner, { with a minimum final signal-to-noise of  $S/N \gtrsim 7$~\AA{}$^{-1}$ being required for consideration further analysis}.
 We display the spectra in the Ca\,{\sc ii} triplet spectral region of the central 
extraction for each of the galaxies in Fig.~\ref{fig:spec}.

\vskip 0.2cm
SDSS images are available for all of the targets.
We used the background subtracted, calibrated images (see Fig.~\ref{fig:images}) of data release DR12 \citep{2015ApJS..219...12A}
for the photometry and other imaging based measurements (see next Section).

\section{Analysis}
\label{section:analysis}

\subsection{Imaging based quantities}
For our analysis we need measurements of the brightnesses, colours, sizes, and ellipticities of the galaxies.
These were obtained by analysing SDSS images as provided by the DR12 pipeline and by querying the SDSS
database. More details are given in the following paragraphs.

\begin{table} 
\begin{center}
\caption{Photometry and other imaging measurements. \label{table:phot}}
\begin{tabular}{lcccc}
\hline \hline
Galaxy           & $M_r$             & $R_{\rm e}$    &  $\epsilon$  &   $g-i$  \\
                 & [mag]             & [arcsec]      & &  [mag]                      \\ 
\hline
LEDA 3115955      & $-19.30$ 	   	& 3.9  & 	0.39  & 1.03 \\   
2MASX J03190758   & $-18.54 $	   	& 5.2  &	0.42  & 1.11 \\   
LCSBS1123P       & $-16.48 $	   	& 7.4  &	0.25  & 0.98  \\  
2MASX J08192430   & $-19.00 	$   	& 13.2 &	0.47  & 1.09  \\  
VIIIZw040        & $-18.72 	   $	& 2.6  & 	0.07  & 1.12 \\   
CGCG038-085      & $-19.17$ 	   	& 7.0  &	0.10  & 1.05 \\   
2MASX J11521124   & $-18.76 	 $  	& 3.2  &        0.17  & 1.02 \\   
CGCG101-026      & $-19.00$ 	   	& 4.4  &        0.29  & 0.99  \\  
LEDA 2108986      & $-18.17$ 	   	& 3.4  &        0.11  & 1.04   \\ 
\hline
\end{tabular}
\parbox{0.45\textwidth}{{\bf Notes: }   The absolute magnitudes  and half-light radii are listed in the  second and third columns. The ellipticities (fourth column) are calculated from the SDSS (DR12, \citealt{2015ApJS..219...12A}) adaptive image moments in the $i$-band (which has the PSF with smallest FWHM) and corrected for the PSF.  
The $g-i$ colours in the fifth column are 7\arcsec\ aperture colours  from the SDSS corrected for Galactic reddening. 
}\end{center}
\end{table}

For the photometric and size measurements in the $r$-band images, we followed an approach similar to \citet{2008ApJ...689L..25J}.
First, the elliptical Petrosian aperture, i.e.~the ellipse at which the isophotal intensity drops to 0.2 times the average intensity within that aperture, was determined. Second, the flux, and the semi-major axis containing half
that flux ($a_{50}$, for which we use the notation $R_\textrm{e}$ hereafter), were measured within an aperture with twice the Petrosian semi-major axis.
Those values were corrected for flux missed by this aperture using concentration measures and the
formulae provided in  \citet{2005AJ....130.1535G}. 
The fluxes were corrected for Galactic extinction \citep[as provided by the SDSS database]{1998ApJ...500..525S},
and converted to absolute magnitudes based on the distances determined as detailed below.

These size measurements  systematically differ from those (circular) Petrosian $R_\textrm{50}$ 
listed in the SDSS database. In order to test the reliability, and to  verify our  sizes, we
also fit S\'ersic models to the galaxies with {\sc galfit} \citep{2010AJ....139.2097P} using models for the
point-spread function (PSF) created with  {\sc SExtractor} \citep{1996A&AS..117..393B} and  {\sc PSFex} \citep{2011ASPC..442..435B}.
 The non-parametric
size measurements and the half-light radii from the S\'ersic model fits are consistent with each other with an average deviation of 6\%
(when disregarding  two galaxies,  which were not well fit with this model, resulting in $n>4$; e.g.~because of a two-component structure). For the further analysis the non-parametric measurements are adopted.

The ellipticities are calculated from the adaptive image moments in the $i$-band (which has the smallest FWHM),
provided by the SDSS pipeline, taking into account the PSF   \citep[see][]{2002AJ....123..583B}.
These ellipticities agree with those from the S\'ersic model fits to the $r$-band images (also taking the
PSF into account) within 12\%, 
(excluding one further outlier, 2MASX~J03190758, for which the deviation is due to radial variations of the ellipticity). 
{ Both ellipticity measurements are light-weighted so that they are a proxy for the  light-weighted average 
 ellipticity within $R_\textrm{e}$.}
For the ellipticities we also adopt the non-parametric measurements. 
 { We also ran the IRAF ellipse task to measure the ellipticities {\it at} $R_\textrm{e}$
 for comparison. For all but two galaxies these ellipticities were compatible within the uncertainties
 with those that we adopted.\footnote{{ The exceptions LEDA 3115955  and 
 2MASX J08192430 (with ellipticities {\it at} $R_\textrm{e}$ of $\sim$$0.29$ and 0.61, respectively) are
  shifted in the $v/\sigma$ and $\lambda$ versus ellipticity diagrams in Section~\ref{section:results}
 when these ellipticities  are used, but this does not change the conclusions. We also note that 
 the comparison sample uses  the average ellipticities {\it within} $R_\textrm{e}$.}}}

Finally, $g-i$ colours are obtained for the sake of a comparison with the sample of Virgo early-type dwarfs. 
For that we queried the SDSS catalog for 7\arcsec~aperture colours and corrected them for Galactic reddening.
The measurements described in this Section are summarised in Table~\ref{table:phot}.

\subsection{Kinematics}
\label{sec:kins}
The recession velocity and velocity dispersion are determined with  \textsc{pPXF} \citep{2004PASP..116..138C}.
The code simultaneously  determines the velocity shift and broadening by velocity dispersion, while
finding  the best-fitting template as a linear combination of the spectra in the template library by using a penalised pixel-fitting method.
In our case, the library consists of velocity standard stars observed with the same setup.

For the fitting two spectral ranges in different echelle orders with a number of metal absorption features are used.
These were the ranges containing the Mg$b$ and Ca\,{\sc ii} triplets ($5100$ \AA{} $< \lambda < 5400$~\AA{} and $8480$~\AA{} $< \lambda < 8710$~\AA{} in the rest frame), both of which also contain several weaker iron absorption lines. In order to obtain estimates for the uncertainties we ran  Monte Carlo simulations of the fitting process by altering the observed spectra by addition of Gaussian noise according to their pixel uncertainties. The final values for the velocities and the central velocity dispersions are the weighted averages (with the inverse variance of the individual values) from the fits in the two independently fit spectral regions. The uncertainties of these values take into account those of the individual measurements as well as discrepancies between them by quadratically adding them (for the velocities: in Table \ref{table:spec} the difference between the Mg and Ca, while in the rotation curves just the difference of the shifted values). The spectra and \textsc{pPXF} fits for the central apertures for the Ca\,{\sc ii} triplet are displayed in Fig.~\ref{fig:spec}.

\begin{figure}  
   \centering
   \includegraphics[height=0.45\textwidth,angle=-90]{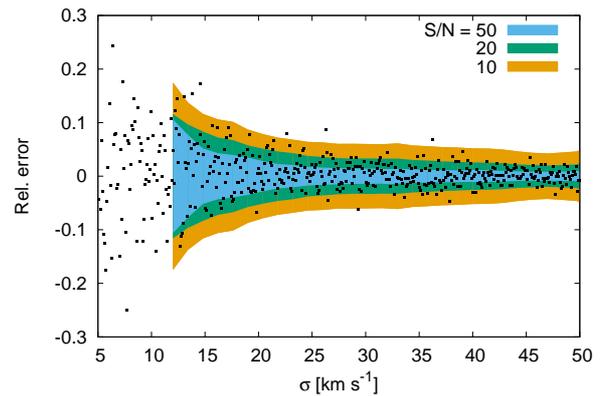} 
      \caption{Relative error versus input $\sigma$ for the Monte Carlo simulation described in Section~\ref{sec:kins}. The contours are the 1-$\sigma$ scatter for the results of the $S/N$  = 50 (blue), 20 (green), and 10 (orange) per pixel runs, and the black points indicate the results for the  simulation with $S/N = 30$ per pixel. Our central velocity dispersions with $S/N$>10 have uncertainties smaller than 10\%.}
   \label{fig:SNsim}
\end{figure}

In all galaxies we required the extracted spectrum of each aperture to have 
$S/N \gtrsim 7$~\AA{}$^{-1}$ for the velocity information to be considered in the further analysis. Furthermore,
the two independently fit spectral ranges had to yield consistent results.
Small systematic offsets in the velocities between the two spectral ranges
were subtracted by determining the offset from the weighted average for each of them
in the central aperture with the highest $S/N$.
The central aperture was also used to derive the central velocity dispersion $\sigma_\textrm{cen}$. 
\citet{Geha:2002kr} found that it is feasible to reliably measure (with an accuracy of 1\%) velocity 
dispersions at the instrumental resolution with a similar setup and $S/N=10$\,pix$^{-1}$
(and with an accuracy of 10\% down to $\sigma\sim18.5$\,km\,s$^{-1}$).

{ We qualitatively confirm this $\sigma$ limit for our ESI instrumental setup with our own idealised Monte Carlo simulation to determine the minimum velocity dispersion that can be recovered by \textsc{pPXF}. 
One of our spectroscopic standard stars, HR1015, which was observed using an identical instrumental setup to our target galaxies, is used as a mock galaxy spectrum for this simulation. 
To simulate an observed galaxy, this simulation takes this high $S/N$ standard star spectrum and over-samples it by a factor of 10. 
This oversampled template is then convolved with 1000 different values of $\sigma_{\rm test}$ between 5~km~s$^{-1}$ and 100~km~s$^{-1}$, to simulate galaxies with different velocity dispersion.
 The simulation then adds noise to the oversampled spectrum until the required $S/N$ per original pixel is met. 
 This  broadened, noise added, normally sampled spectrum is then used as a mock galaxy spectrum, with the original, high $S/N$, un-broadened spectrum used as a template by \textsc{pPXF} (i.e.~the simulation does not take into account possibly template mismatches).
\textsc{pPXF} is then run to test how well the artificial velocity broadening was recovered. The results of this simulation are shown in Fig.~\ref{fig:SNsim}  for four different $S/N$ ratios (10, 20, 30, and 50 per pixel). }

The spectra in the central apertures have at least $S/N \ge10$\,pix$^{-1}$, in both spectral ranges (in most cases larger by some),
for all our observations, and we note that all the measured central velocity dispersions exceed the instrumental resolution ($\sigma_\textrm{instr}\sim25$\,km\,s$^{-1}$), and are thus reliable. 
We also note that the central velocity dispersions are consistent within the errors with those listed in the SDSS database
for all galaxies but those with the most significant rotation (see below). The higher values in SDSS in these cases reflect
the larger spatial coverage of the SDSS fibre ($3\arcsec$ and 2\arcsec\ for the SDSS and BOSS spectrographs, respectively).

\begin{table}
\begin{center}
\caption{Spectroscopic measurements. \label{table:spec}}
\scalebox{0.85}{
\begin{tabular}{lccccc}
\hline \hline 
Galaxy           & $V$      & $D$   & $\sigma_{\rm cen}$  & $V_{\rm rot,e}$ & $\left<\sigma\right>_{\rm e}$      \\
                 & [km\,s$^{-1}$] &  [Mpc]    & [km\,s$^{-1}$] & [km\,s$^{-1}$]   & [km\,s$^{-1}$]        \\
\hline
LEDA 3115955          & $5808\pm 11 $ &    83.0 & $88\pm 1  $ & $27\pm2$     &  $89\pm1$ \\  
2MASX J03190758   & $4076\pm 27 $ &    58.2 & $64\pm2   $ & $19\pm3$     &  $64\pm2$ \\  
LCSBS1123P            & $1905 \pm 7 $&    27.2 & $28\pm 1    $ & $5 \pm1$     &  $29\pm1$ \\ 
2MASX J08192430   & $3894\pm 14 $ &    55.6 & $51\pm1   $ & $69\pm4$     &  $60\pm2$ \\ 
VIIIZw040                  & $3662\pm  13$ &    52.3 & $99\pm 3  $ & $1\pm3$      &  $99\pm3$ \\  
CGCG038-085          & $3692\pm 8 $ &    52.8 & $94\pm 1    $ & $74\pm13$ & $100\pm3$ \\  
2MASX J11521124    & $5270\pm  13$ &    75.3 & $69\pm 1  $ & $19\pm3$     &  $70\pm1$ \\  
CGCG101-026          & $3627\pm 10 $ &    51.8 & $48\pm 3  $ & $49\pm3$     &  $53\pm3$ \\ 
LEDA 2108986          & $2546\pm 4 $ &    36.4 & $36\pm 2    $ & $8\pm4$  &  $37\pm2$ \\
\hline
\end{tabular}
}
\parbox{0.45\textwidth}{{\bf Notes:   } The heliocentric recession velocities (second column) are used to calculate the distances in the third column,  assuming a Hubble flow with $H_0 = 70$\,km\,s$^{-1}$\,Mpc$^{-1}$. The recession velocities and central velocity dispersions (fourth column) are weighted means of the fits to the Ca\,{\sc ii} and Mg$b$ triplet spectral regions, the uncertainties give the statistical uncertainties (see text). The fifth column
lists the rotation velocity at $R_\textrm{e}$ of the \textit{Polyex} model { with the asymptotic standard error estimate from the Levenberg-Marquardt fitting algorithm} (for LEDA 2108986 the alternative model fit, shown with a green curve in Fig.~\ref{fig:rotcurves}, results in $V_{\rm rot,e} = 18\pm1$ km\,s$^{-1}$).
The final column lists the integrated   $\left<\sigma\right>_{\rm e}$ within $R_\textrm{e}$, assuming the \textit{Polyex} rotation curve, a constant $\sigma$, and a S\'ersic index of the surface brightness profile of $n=2$ { along with the propagated uncertainties from $V_{\rm rot,e}$ and $\sigma_{\rm cen}$} (see text).
}
\end{center}
\end{table}

\subsection{Harmonisation of kinematics}
A comparison of the rotation curves for our sample,
and ultimately with the kinematics of early-type dwarfs
in the Virgo cluster, needs to be carried out in a consistent way.
The rotation curves, e.g.\ for  exponential discs, are 
expected to be rising up to 2.2  scale lengths \citep{1970ApJ...160..811F}, 
i.e.\ beyond  the typical extent of our measurements and
those in the literature. Therefore, the rotation amplitude at
a common radius is used. We chose the galaxy half-light radius $R_\textrm{e}$,
as also done by \citet{2015ApJ...799..172T} for the Virgo dwarfs.

\begin{figure*}  
   \centering
   \includegraphics[height=0.33\textwidth,angle=-90]{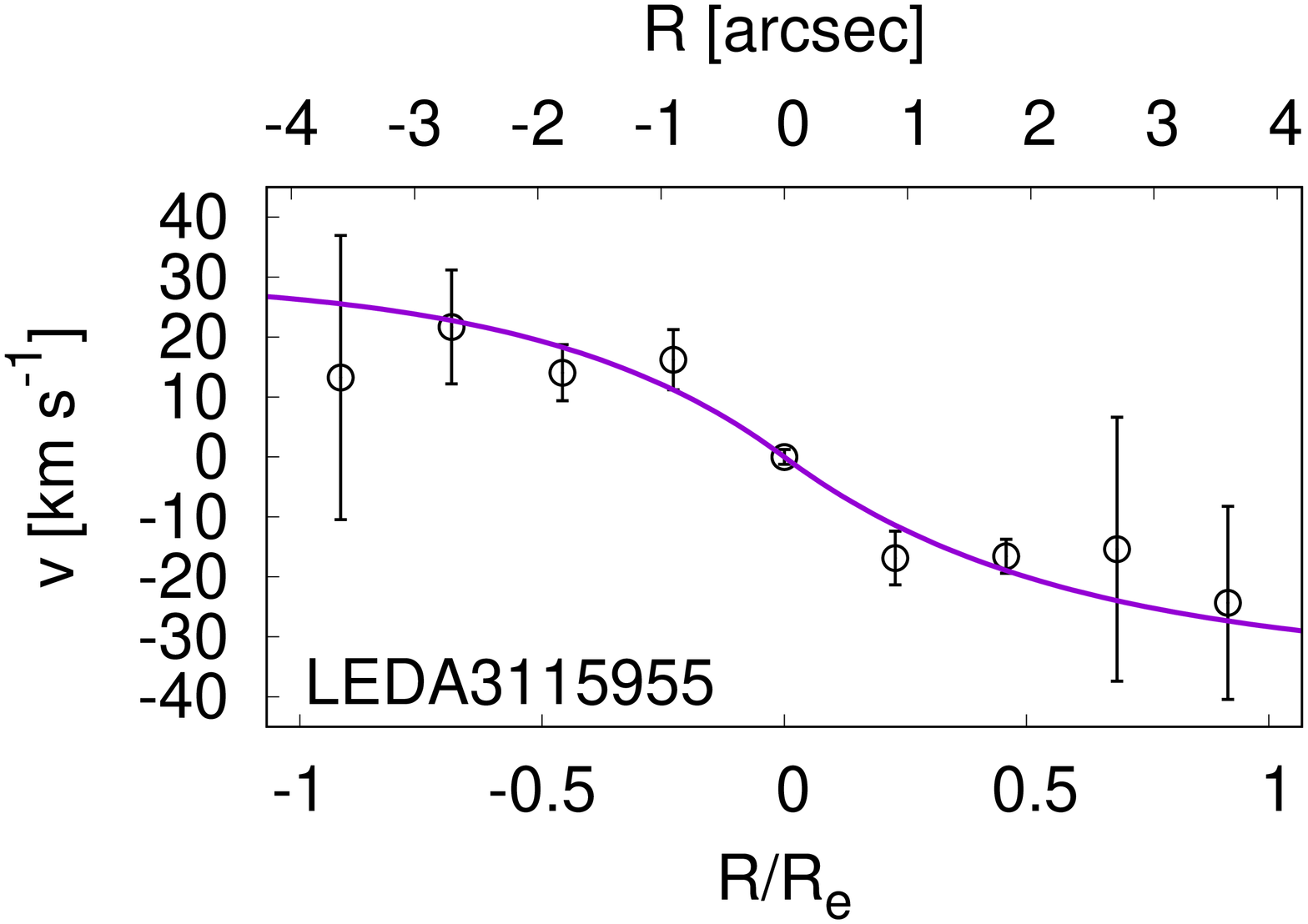} 
   \includegraphics[height=0.33\textwidth,angle=-90]{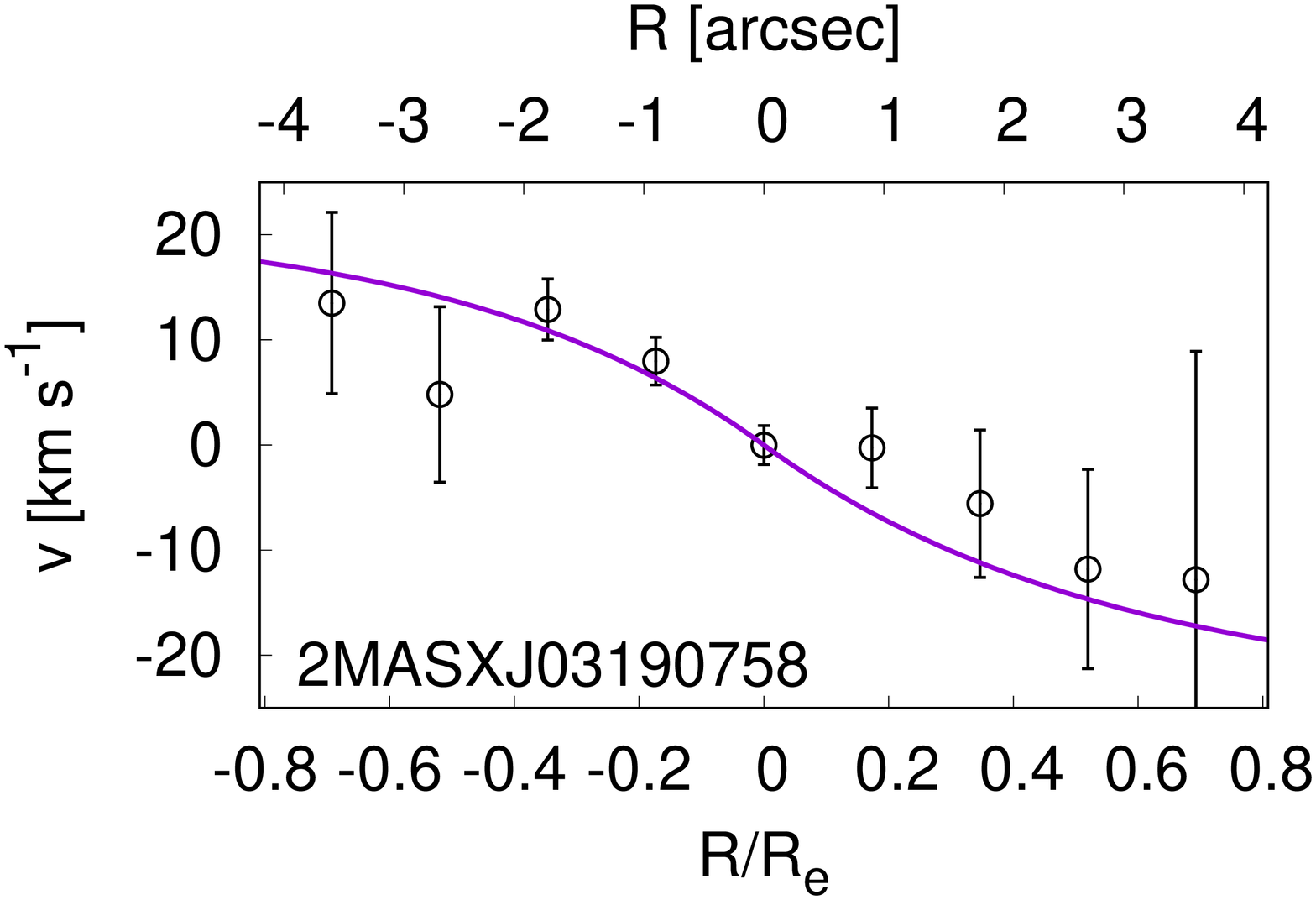} 
   \includegraphics[height=0.33\textwidth,angle=-90]{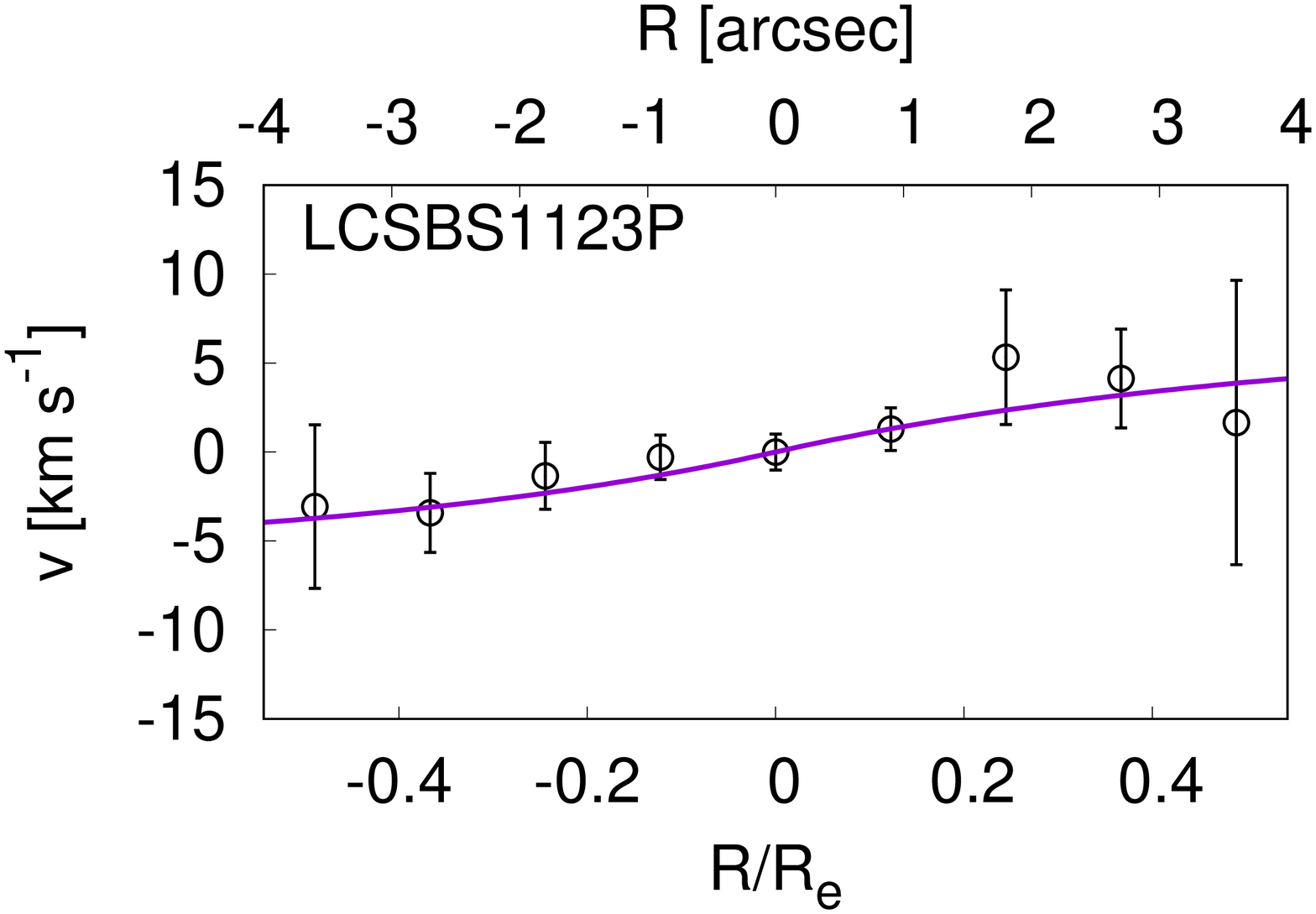} \\
   \includegraphics[height=0.33\textwidth,angle=-90]{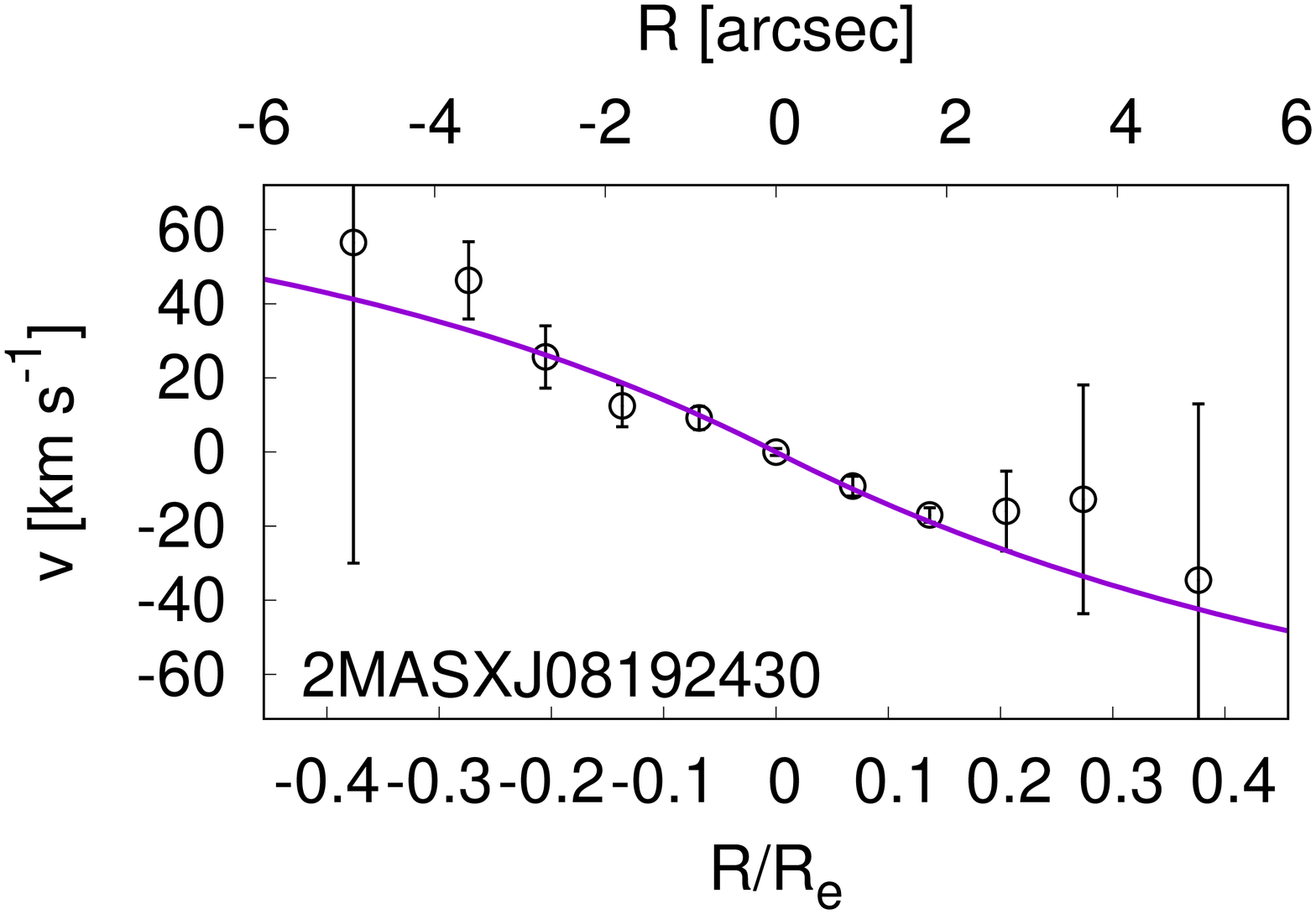} 
   \includegraphics[height=0.33\textwidth,angle=-90]{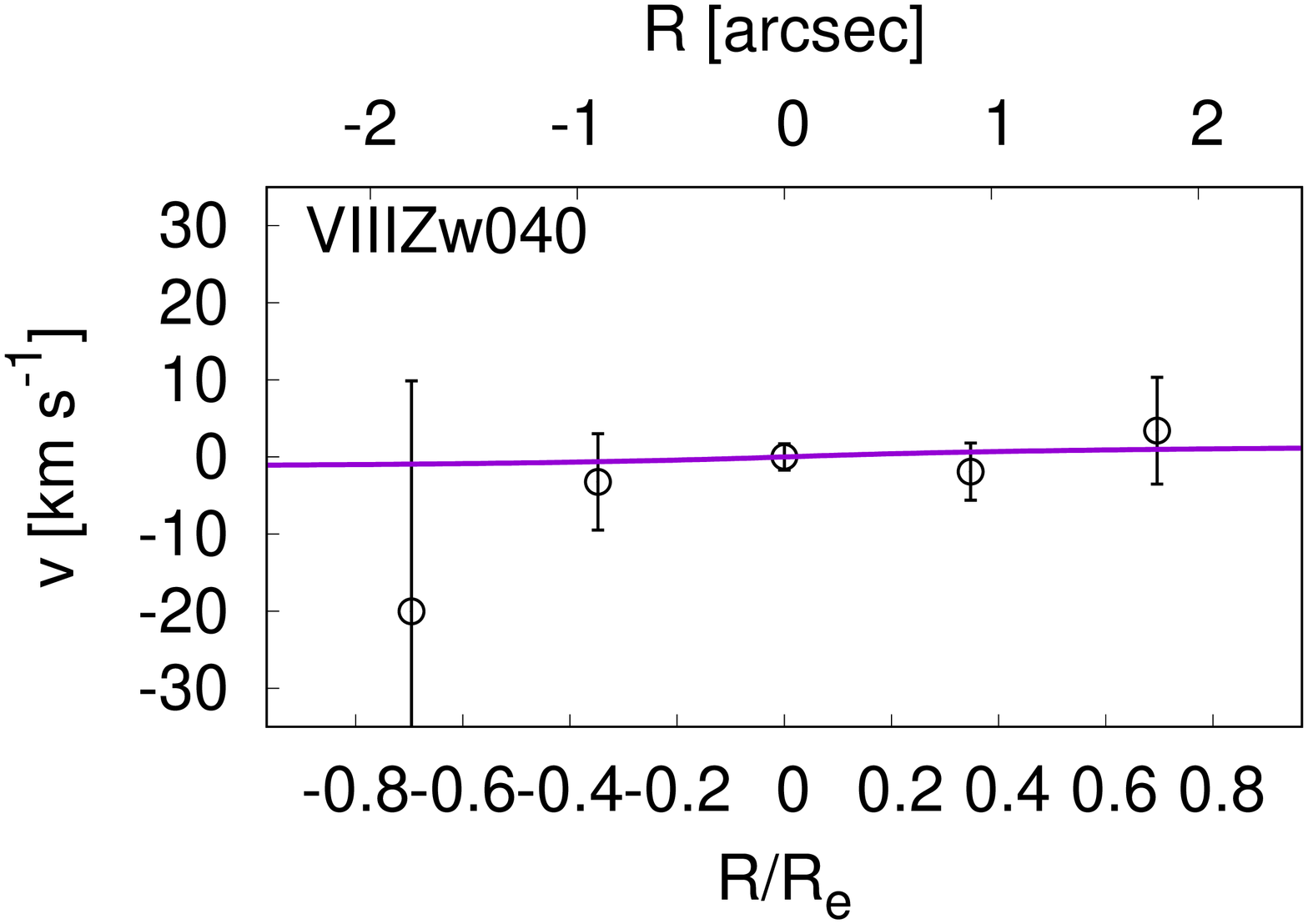} 
   \includegraphics[height=0.33\textwidth,angle=-90]{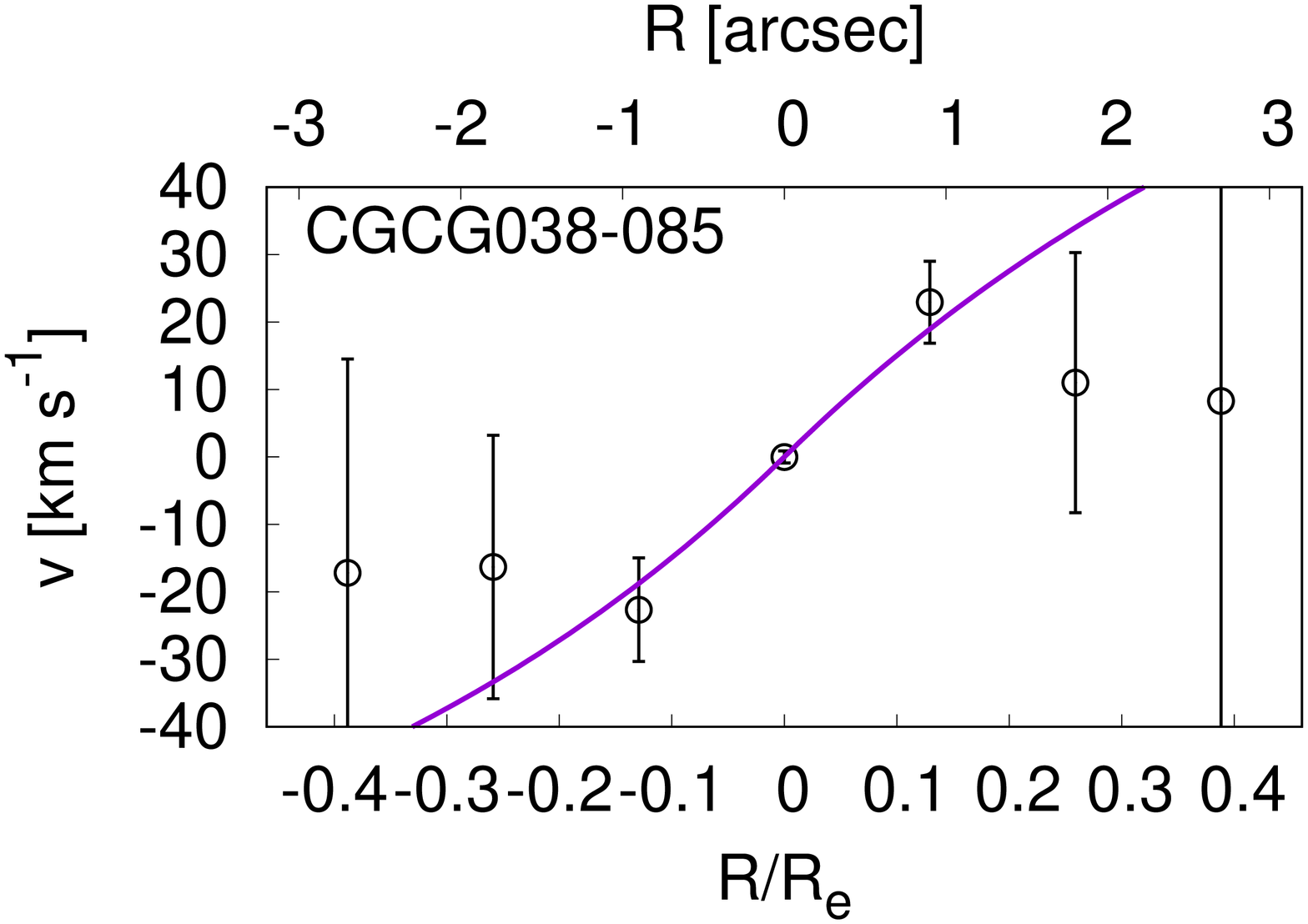} \\
   \includegraphics[height=0.33\textwidth,angle=-90]{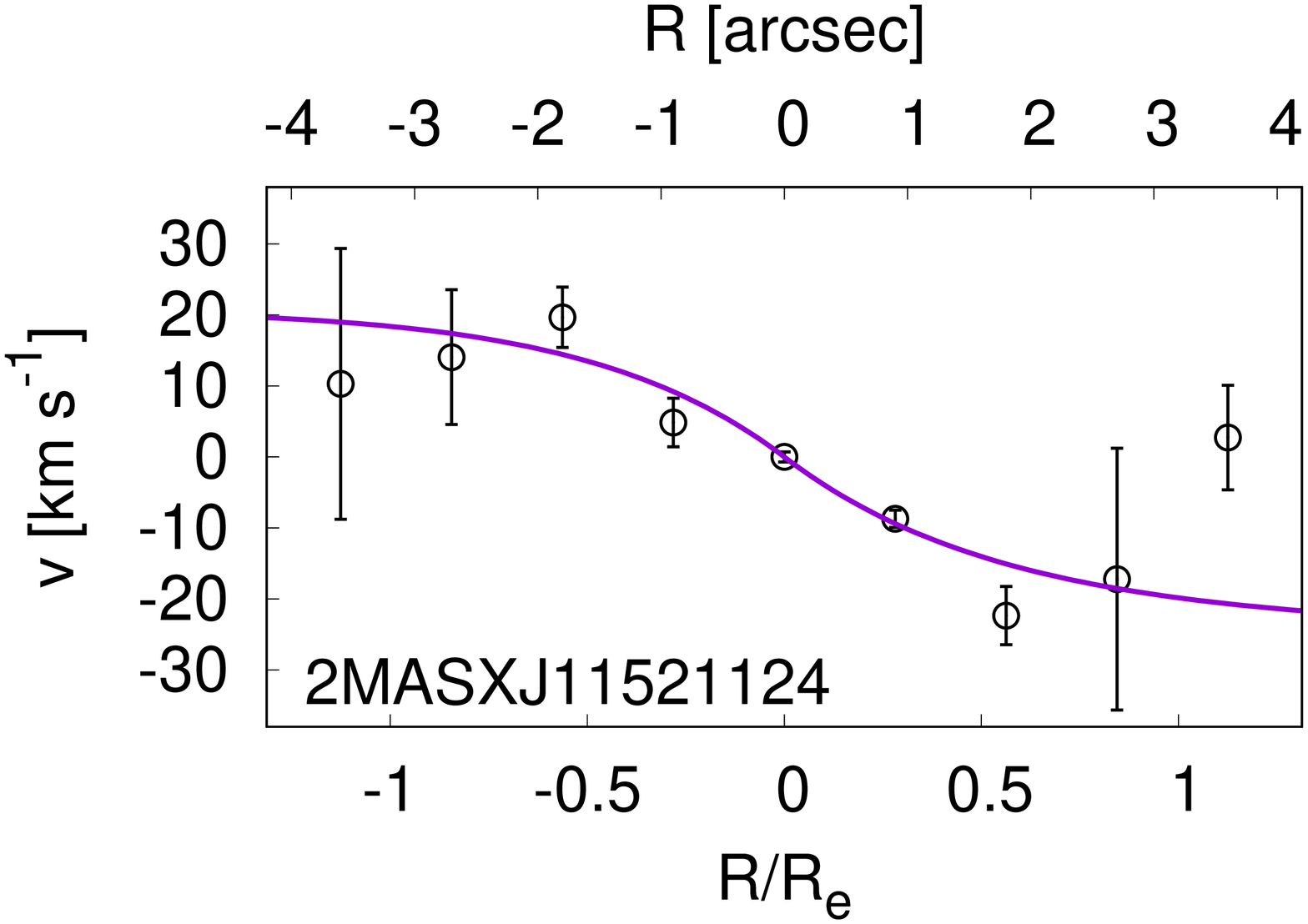} 
   \includegraphics[height=0.33\textwidth,angle=-90]{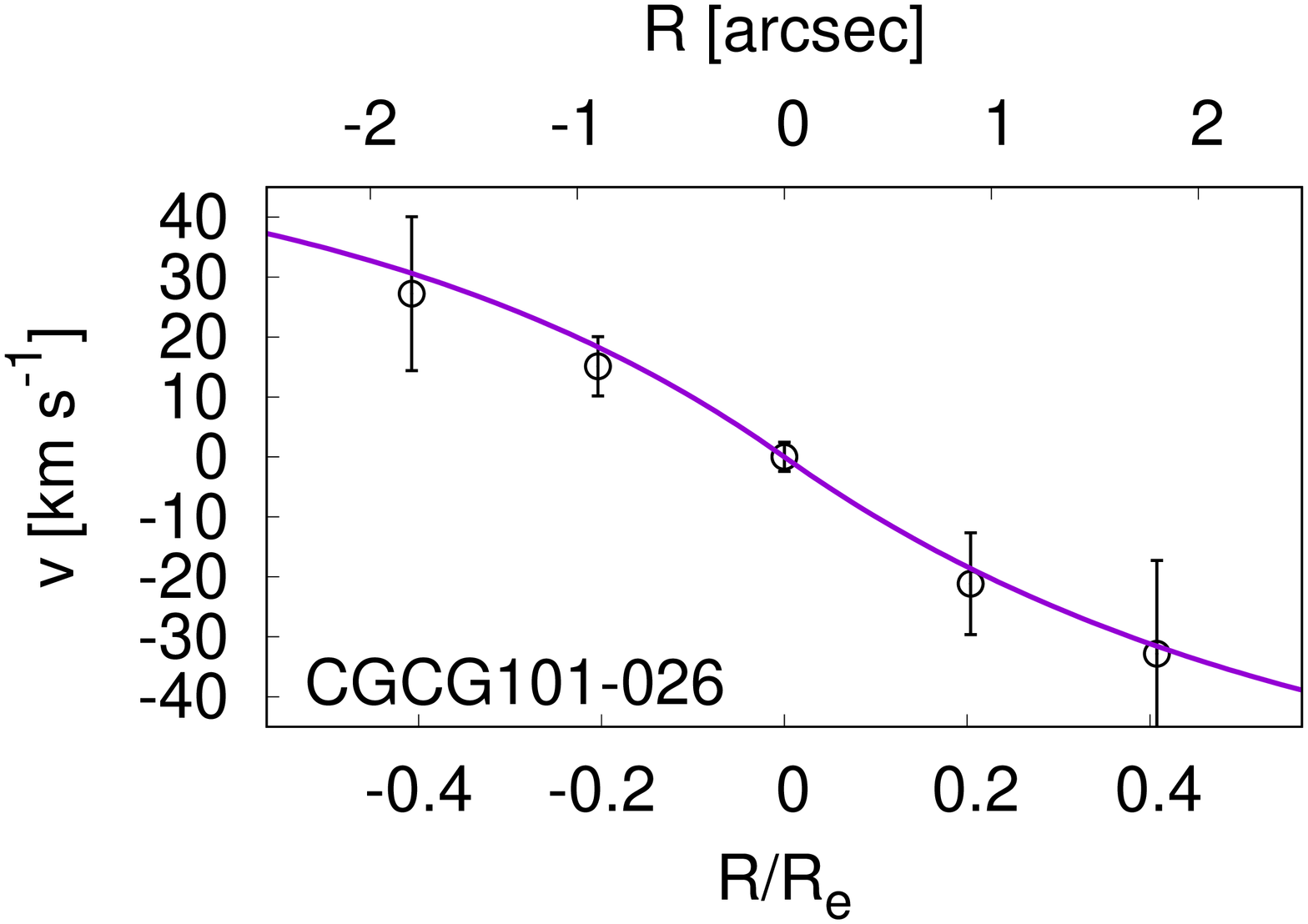} 
   \includegraphics[height=0.33\textwidth,angle=-90]{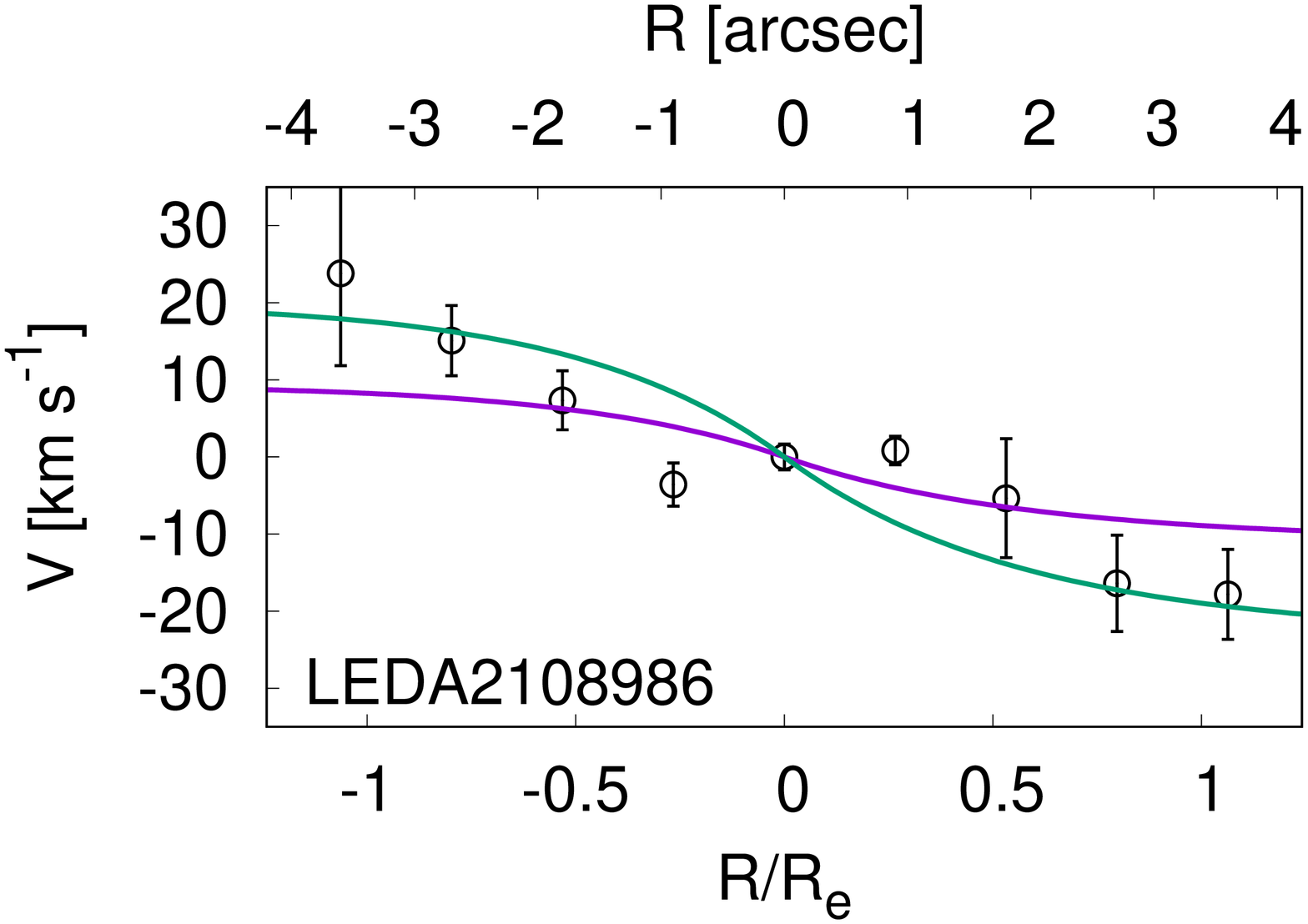} 
      \caption{Rotation curves of galaxies in our sample. The uncertainties are from Monte Carlo simulations of the measurements and additionally take into account the consistency of the independent measurements for the two different spectral ranges used for fitting. The curves display fits of the {\it Polyex} model to the rotation curves (see text). For LEDA 2108986 an alternative fit was obtained by disregarding the inner parts with a possibly  distinct kinematic feature, and is shown as green curve.}
   \label{fig:rotcurves}
\end{figure*}

Not all of our rotation curves reach $R_\textrm{e}$, nor do they
uniformly sample the rotation amplitude at that radius, i.e.~the apertures from which the spectra were extracted  do not correspond to equal physical scales for different galaxies. 
For inter- and extrapolation of the rotation curves at $R_\textrm{e}$
we fit model rotation curves (Fig.~\ref{fig:rotcurves}) parametrised by the \textit{Polyex} function
suggested by \citet{2002ApJ...571L.107G}:
\begin{equation}
 V_\textrm{poly}(R) = V_0 \left\{(1- e^{-R/R_\textrm{PE}}) \times (1+\alpha R/R_\textrm{PE})\right\},
 \end{equation}
with the rotation amplitude $V_0$. 
The scale length  of the steep inner rise $R_\textrm{PE}$, and the slope in the slowly varying outer parts $\alpha$
are not well constrained by our data with its radial limitations.
\citet{2006ApJ...640..751C} fit \textit{Polyex} models to a large number of
rotation curves of disc galaxies and co-additions thereof, increasing the radial coverage
to several exponential scale lengths $h$. They concluded for low luminosity galaxies
$\alpha \sim 0.02$  and $R_\textrm{PE} \sim h$. In our fits we use these values as fixed parameters
(with $h\approx0.6\, R_\textrm{e}$ for exponential profiles) leaving only the rotation amplitude as a free parameter.
We note that the \textit{Polyex} model  is a versatile phenomenological model used for rotating discs and that we use the galactic size and we do not try to separate possibly embedded discs.
This approach was also followed by \citet{2015ApJ...799..172T}, who confirmed the
applicability of these approximations for the rotation curves of early-type dwarfs in their sample 
with larger radial extent. For these objects \citeauthor{2015ApJ...799..172T} also compared
the rotation amplitudes obtained from fits to the whole rotation curve with those from fits with
an artificial restriction to $R<0.6\, R_\textrm{e}$. The good agreement suggests that our extrapolations
of the rotation curves to $R_\textrm{e}$ for those objects with less extent is reasonable.

The  velocity dispersion profiles  of early-type dwarf galaxies are generally flat \citep[e.g.][as expected for systems with low S\'ersic $n$, see \citealt{1997MNRAS.287..221G,2005MNRAS.362..197T}]{2003AJ....126.1794G,2009MNRAS.394.1229C,2011MNRAS.413.2665F,2014MNRAS.439..284R,2015ApJ...799..172T,2015MNRAS.453.3635P}.
The objects in our sample for which we can extract radial information of the velocity dispersion confirm this within the uncertainties. For the further analysis we chose to use for all galaxies the velocity dispersion measured in the central
aperture with the highest $S/N$. We  note that this would also be the dominating contributor, if the velocity dispersion
was averaged over larger radial extent when weighting with the luminosity. The observed flat velocity dispersion profiles also suggest that the central value is not severely biased by the potentially different velocity dispersion of a nuclear star cluster (as, e.g., in NGC~205; \citealt{1990MNRAS.245P..12C}; see also \citealt{2011MNRAS.413.2665F}).

When comparing rotation and random motions we adopt a  similar approach to \citet{2015ApJ...799..172T}.
As described above, we quantify the rotation as the amplitude of the best fitting {\it Polyex} model at $R_\textrm{e}$.
The rotation is compared to a luminosity-weighted average of the second moment of the velocity field $\left<\sigma\right>$
(with contributions from both the random motions and the rotation, see below).
\citet{2015ApJ...799..172T} achieved this by measuring the broadening of absorption lines on spectra, which were obtained by luminosity-weighted co-adding the spectra from within $R_\textrm{e}$ on both sides of the galaxy centre.
However, for finding the average broadening along the slit within $-R_\textrm{e}$<$R$<$+R_\textrm{e}$, we are limited by the
radial extent of our observations. Instead of collapsing the spectra within such a large aperture,
we calculated our values $\left<\sigma\right>_{\rm e}$ based on the \textit{Polyex} model fits and the assumption that the velocity dispersion is independent of the radius (see above):
\begin{equation}
 \left<\sigma\right>_\textrm{e}^2 =\int_{-R_\textrm{e}}^{+R_\textrm{e}}  V^2_\textrm{poly}(R) I(R) dR \bigg/ \int_{-R_\textrm{e}}^{+R_\textrm{e}}  I(R) dR
\ \ \ \ \  \ \ + \ \ \sigma_\textrm{cen}^2 
  \end{equation}
 The relative contribution of the rotation to the luminosity-weighted average depends on the surface brightness profile. We numerically carry out the weighted average   assuming different surface brightness profile shapes. For this the surface brightness profile is parametrised by a S\'ersic profile and we consider cases with indices ranging from $n=1$ to 4, a range which easily contains those indices expected for our galaxy sample. We list the averages for $n=2$ and summarise all measurements described in this Section in Table~\ref{table:spec}.

{ Furthermore, we calculate the spin parameter $\lambda_\textrm{e}$ \citep{2007MNRAS.379..401E} in the following way, with

\begin{equation}
\lambda_\textrm{e,1D} = \sum_i F_i R_i | V_i |  \ \ \ \bigg/  \ \ \ \sum_i F_i R_i \sqrt{V_i^2+\sigma_i^2}
\end{equation}

\noindent summing over several apertures along the slit within $R_\textrm{e}$, with the flux in an aperture $F_i$, its radial distance to the galaxy centre $R_i$, and its rotation velocity and velocity dispersion $V_i$ and $\sigma_i$.
Again, due to the unequal sampling of the rotation with small numbers of apertures, we base our $\lambda_\textrm{e}$ calculation
on the model rotation curve, the assumption of a spatially constant velocity dispersion, and a S\'ersic light profile (i.e.~$V_i=V_\textrm{poly}(R_i)$, $\sigma_i=\sigma_\textrm{cen}$, and  $F_i= F_\textrm{e}*\exp\left(-b_n\left[(\frac{R}{R_\textrm{e}})^{\frac{1}{n}}-1\right]\right)$, where the flux at the half-light radius $F_\textrm{e}$ cancels out in the calculation of $\lambda_\textrm{e}$ and we take an lower and upper extreme S\'ersic index $n=1$ and $n=4$ to estimate its influence).
\citet{2015ApJ...799..172T} applied a factor to convert the long-slit information to a value mimicking  a measurement integrated over the whole half-light aperture, corresponding to what integral field spectroscopy would measure: $\lambda_\textrm{e} = 0.64\,\lambda_\textrm{e,1D}$.  
This factor was determined from comparisons of modelled long-slit parameters and corresponding two-dimensional quantities using realistic velocity fields and surface brightness distributions \citep[see][]{2015ApJ...799..172T}. For maximal comparability, we employ the same factor.}

\begin{figure}  
   \centering
   \includegraphics[width=0.48\textwidth,angle=0]{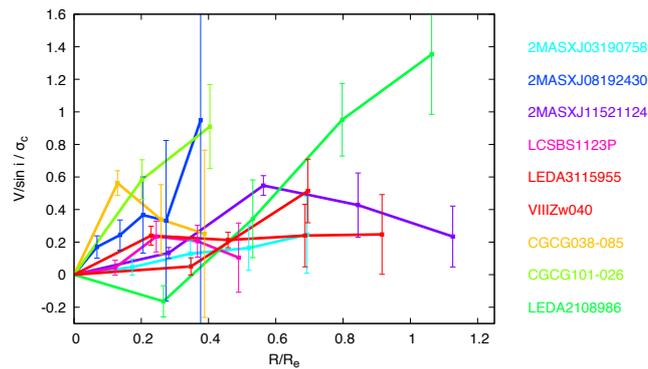} 
      \caption{Rotation curves corrected for inclination (from ellipticity) and normalised by central velocity dispersion. The approaching and receding sides are folded together and averaged. A variety of different types of profiles can be seen.}
   \label{fig:normrotcurves}
\end{figure}

\section{Results}
\label{section:results}

We obtained rotation curves for 9 isolated quenched low-mass galaxies (Fig.~\ref{fig:rotcurves}).
For three galaxies their extent is equal to or exceeds $R_\textrm{e}$. For two galaxies, $\sim0.7\, R_\textrm{e}$
are reached, while for the remaining four $R_\textrm{e}/2$ is covered. In the following sections 
we present the results in more detail and compare them to  those for a sample of Virgo cluster dwarfs.

\subsection{Variety of rotation profiles}

At first we are interested in an internal comparison of the rotation curves within our sample. 
For that we correct the rotation velocity for the inclination, which is estimated by the axis ratio, and
normalise it to the central velocity dispersion, and likewise the radii to $R_\textrm{e}$ (Fig.~\ref{fig:normrotcurves}).
Instead of a clear pattern arising, a range of different behaviours can be seen.
There are rather rapidly rising curves, as well as very flat ones for objects that basically do not rotate.

Since this information is extracted from slit spectra, rotation could be missed when the
rotation is misaligned with the major axis. 
Also, the slit could be misaligned when the position angle varies as a function of radius, or when the object
is very round, which applies particularly to VIIIZw040. 
For this galaxy, however, its  compactness  and especially its high velocity dispersion suggest
that it is genuinely dominated by random motions.

The profiles of CGCG038-085, and possibly 2MASX~J11521124, are somewhat reminiscent of 
rotation curves of galaxies with bars \citep[e.g.][and references therein]{1987MNRAS.225..653S,1989AJ.....97...79B}.
For CGCG038-085, there is a slight hint of a bar-like structure in the S\'ersic model subtracted 
residual image.
{  For LEDA 2108986 the innermost part of the rotation profile appears to be flat. 
This  may hint at a kinematically decoupled core (KDC) with counter-rotating stars with respect
to the outer motions. This claim is substantiated by counter-rotating ionised gas 
 (rotating in alignment with a possible counter-rotating inner structure), which is discussed, along with an inner spiral structure, in a separate paper (Graham et al.~2016).}

\subsection{Comparison with Virgo cluster early-type dwarfs}
\begin{figure}  
   \centering
   \includegraphics[height=0.5\textwidth,angle=-90]{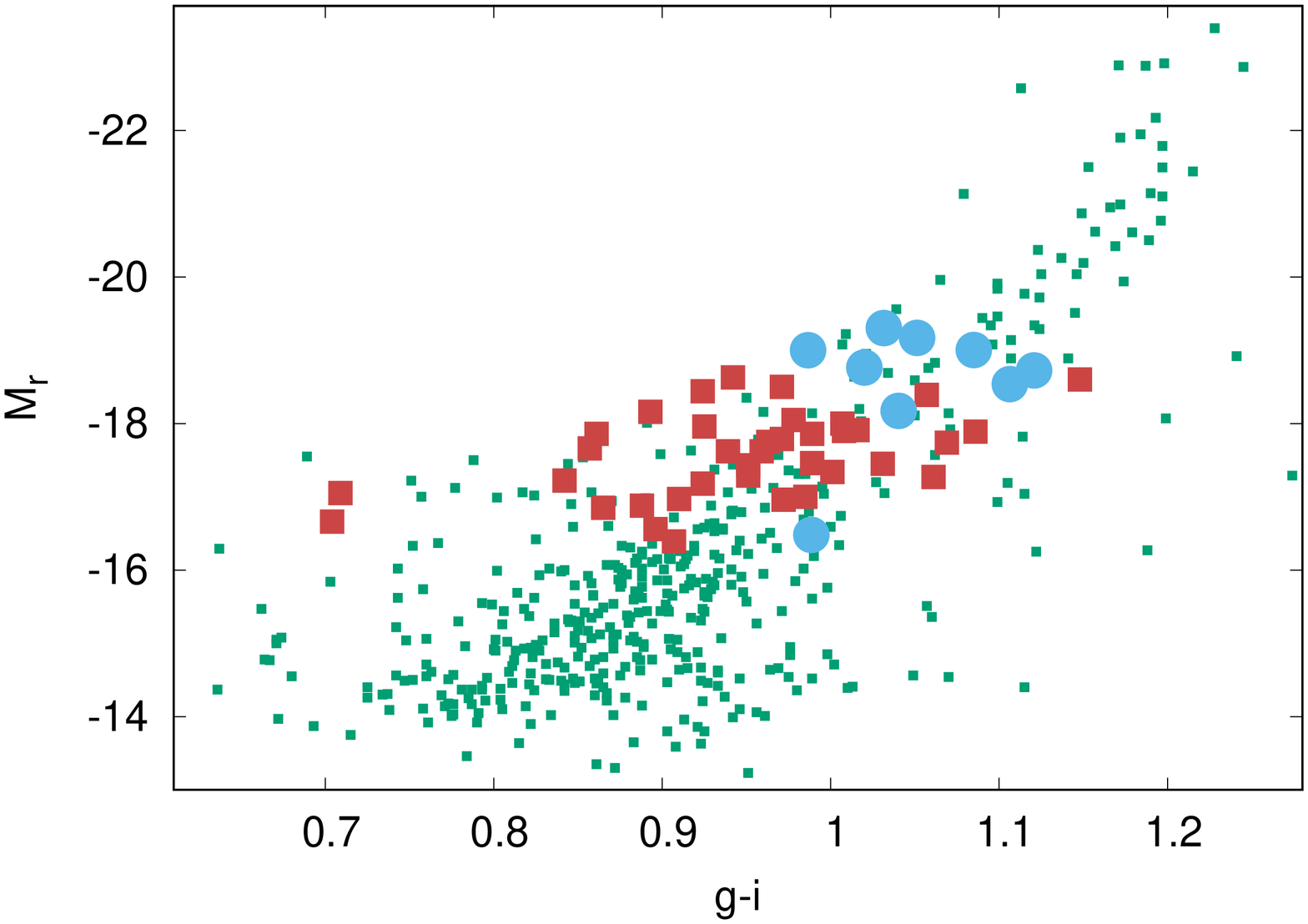} 
   \includegraphics[height=0.5\textwidth,angle=-90]{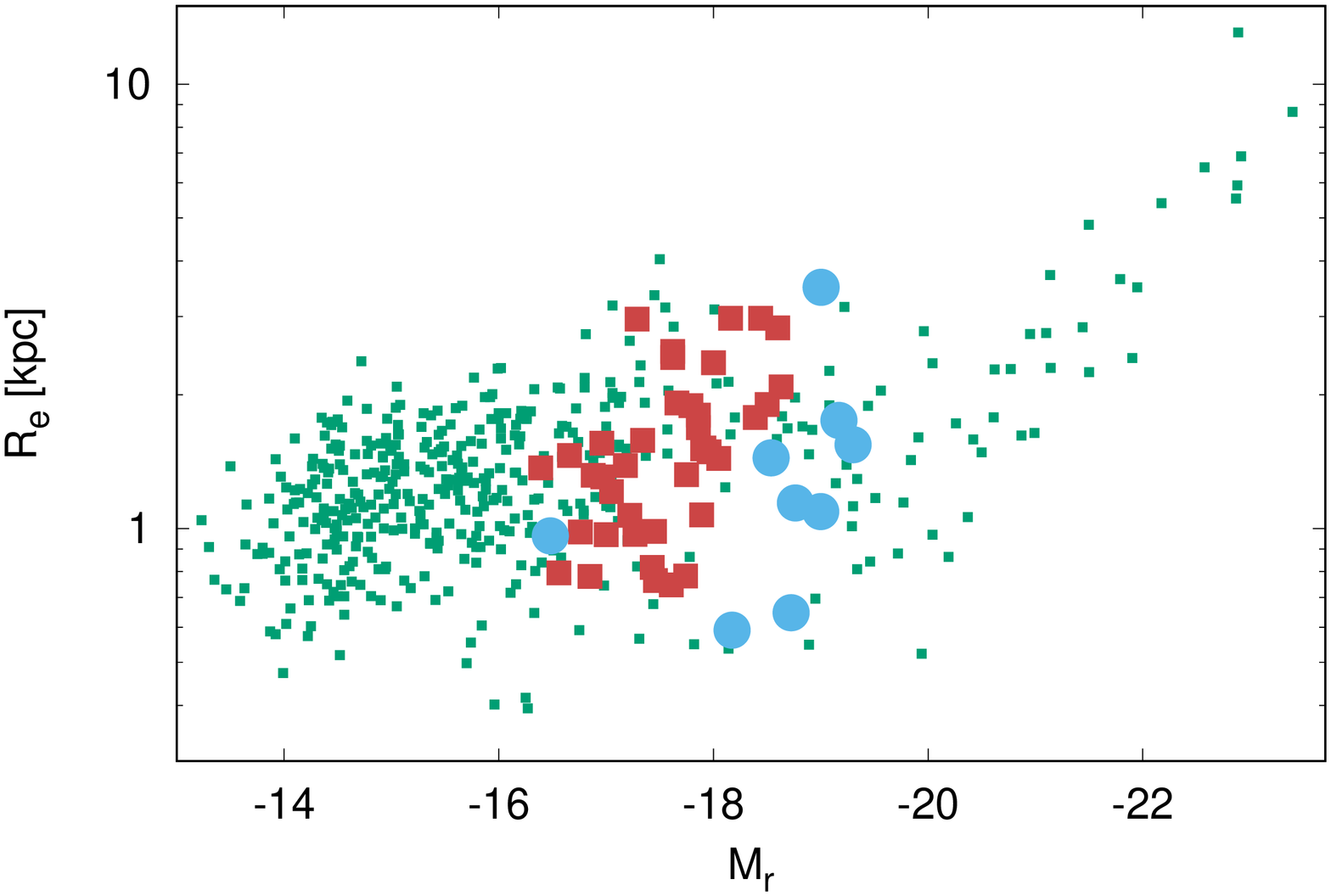} 
      \caption{Colour magnitude and size magnitude relations. Shown are the Virgo early types (green dots; from \citealt{2008ApJ...689L..25J,2009ApJ...696L.102J}), the sample of \citet[red squares]{2014ApJS..215...17T}
      and our galaxies (blue points --  extinction corrected 7\arcsec\ aperture colours). This comparison shows that the isolated quenched galaxies  fall on the red sequence of the Virgo early types and span a very similar range of sizes. }
   \label{fig:cmr}
\end{figure}

Before comparing the internal kinematics of our galaxy sample to those of early-type dwarfs in the Virgo cluster, we consider other basic properties,
i.e.~colour magnitude and size magnitude relations (Fig.~\ref{fig:cmr}). 
The galaxies in our sample 
have on average somewhat brighter absolute
 $r$-band magnitudes compared to those in the \citet{2014ApJS..215...17T} comparison sample. 
 { This is also reflected in the velocity dispersions, which are larger than
 the largest dispersion in the  \citet{2014ApJS..215...17T} sample for five of the galaxies in our sample.}
The faintest galaxy in both samples are of similar absolute brightness.
Consistent with the selection as quenched galaxies, the galaxies in  our sample fall on the red sequence
traced by the Virgo early types \citep{2009ApJ...696L.102J}.
It is  noteworthy that their sizes span most of the size range of the Virgo early-type dwarfs \citep{2008ApJ...689L..25J},
and their range of sizes exceeds that of the \citet{2014ApJS..215...17T}  sample.

\begin{figure}  
   \centering
   \includegraphics[height=0.5\textwidth,angle=-90]{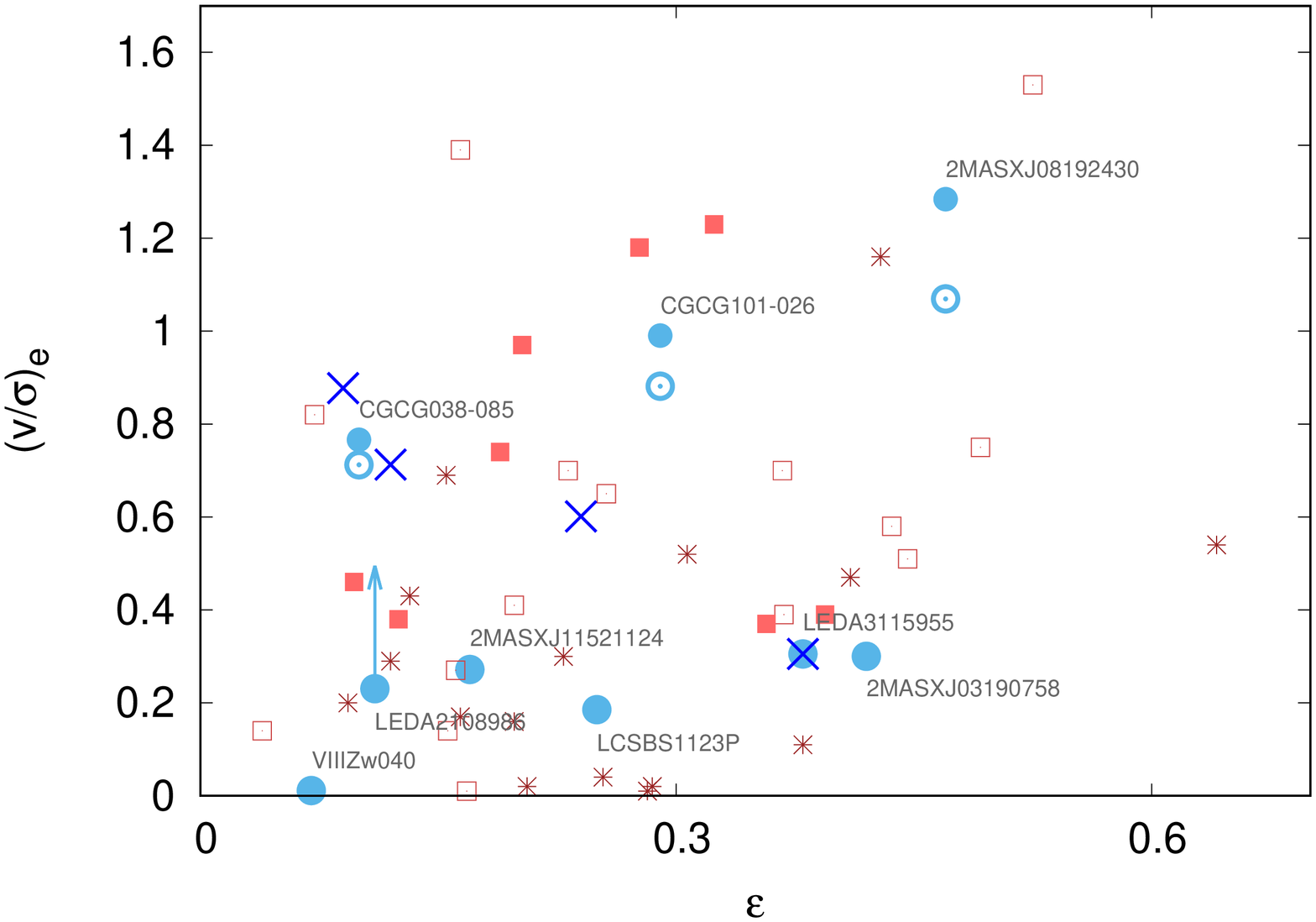} 
   \includegraphics[height=0.5\textwidth,angle=-90]{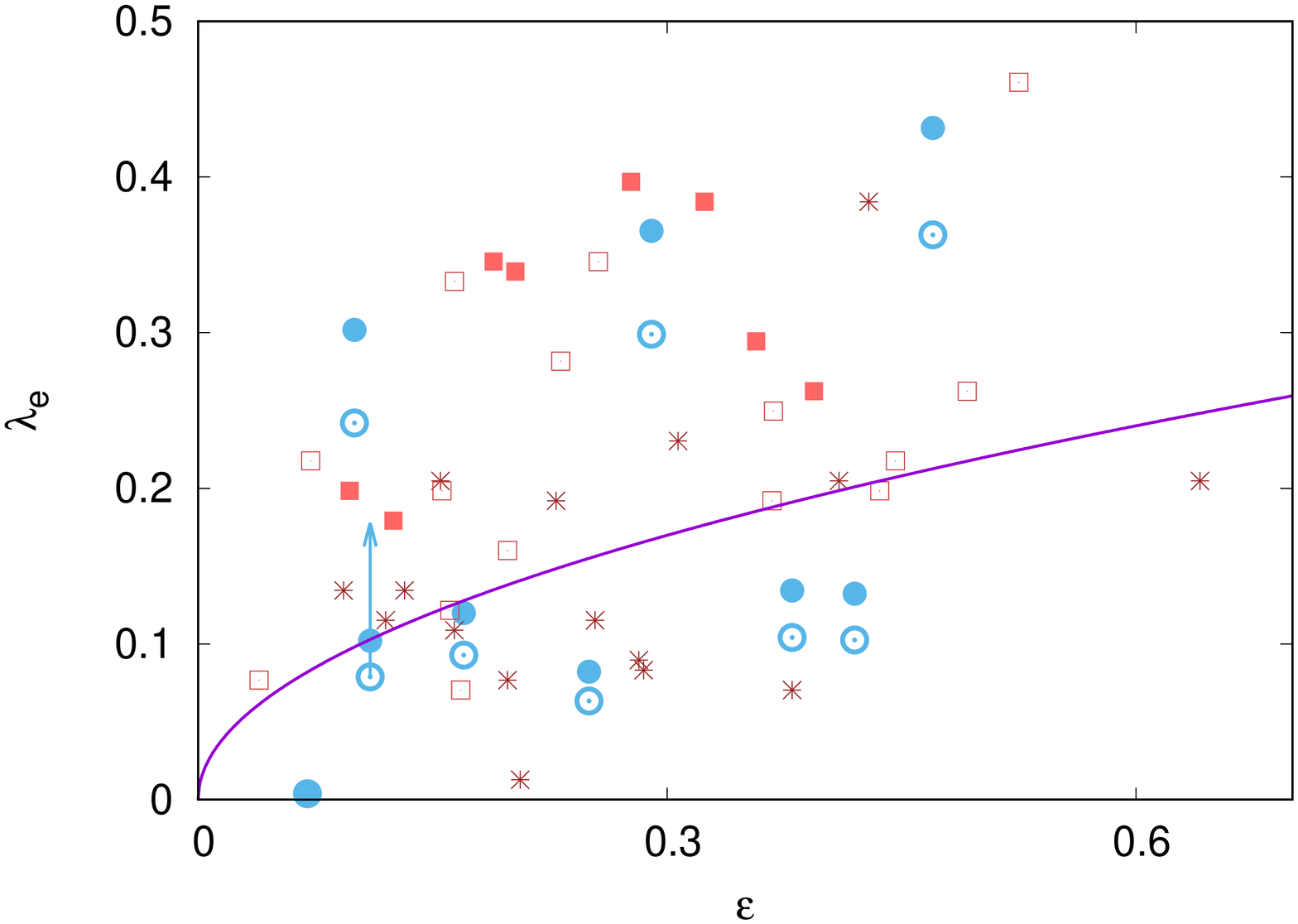} 
      \caption{ { Upper panel: Rotation parameter $(v/\sigma)_\textrm{e}$  as function of ellipticity. Light blue open and filled symbols display the calculations for our sample galaxies assuming a surface brightness profile with S\'ersic index $n=1$ and $4$, respectively.
      The arrow shows the shift for LEDA 2108986, when considering the alternative fit (see text).
     The different red symbols show their Virgo early-type dwarfs (the boxes, circles, and dots indicating small ($D$<2$^{\circ}$=0.4 $R_\textrm{vir}$, \citealt{2012ApJS..200....4F}), intermediate, and large projected distances ($D$>$R_\textrm{vir}$) to the cluster centre, respectively).  The blue crosses indicate the stellar kinematics of BCDs from  \citet{2014MNRAS.441..452K}.
     Lower panel: $\lambda_{\textrm{e}}$ vs ellipticity with the same symbols.
       The purple line marks the demarcation line between slow and fast rotators from \citet{2015ApJ...799..172T}.
     There is no clear cut difference in the kinematics of isolated early-type low-mass galaxies when compared to those in the Virgo cluster, and the four BCDs of \citet{2014MNRAS.441..452K} share the same parameter space.}}
   \label{fig:vsigma}
\end{figure}

\citet{2015ApJ...799..172T} found a similar variety of rotation profiles
with different slopes and amplitudes for early-type dwarf galaxies in the
Virgo cluster.
The distribution of our galaxies in  the $(v/\sigma)_\textrm{e}$ and $\lambda_\textrm{e}$ versus ellipticity diagrams
(Fig.~\ref{fig:vsigma}) is not very different from that of the Virgo dwarfs as
a whole population.
\citeauthor{2015ApJ...799..172T} found 11 slow versus 28 fast rotators,
while our sample has around half of the galaxies in either class (Fig.~\ref{fig:vsigma}). 
\citet{2014ApJ...783..120T} concluded that $\sim$6\% of early-type dwarfs have KDCs \citep[see also][]{2016MNRAS.462.3955P}.
One KDC in our sample of nine galaxies { would be} consistent with this frequency  given the small
number statistics.

\section{Discussion}
\label{section:discussion}

\subsection{Environmental dependencies and isolation}
\label{sec:iso}

One line of argument in favour of the `galaxy harassment in clusters' scenario is
the identification of a trend in some characteristic as a function of a quantification
of the environment. For example, \citet{2015ApJ...799..172T} found an increase of the spin parameter in the dwarfs in the Virgo cluster as a function of the 
(projected) distance to the cluster centre \citep[cf.~Fig.~\ref{fig:vsigma}; see, however,][who found with a smaller sample that such a trend is not significant when using 3D distances instead of projected ones]{2013MNRAS.428.2980R},
 in this case given by the location of M87.
Such a correlation of  the spin parameter with  clustercentric distance
would be consistent with the finding that
 early-type dwarfs with signs of discs in their morphologies \citep{2006AJ....132..497L} 
 are less clustered towards the centre than those without. 
Our isolated objects do not follow an extrapolation of such a trend, but  instead show a similar variety of kinematic configurations as those in the
Virgo cluster. We note  that the analysis of dwarfs in galaxy groups by \citet{2015MNRAS.453.3635P} 
did not find any evidence in the kinematic scaling relations that would suggest 
a continuous change from the cluster centre to its outskirts to less rich groups and the field. { We also note that  higher mass early-type fast rotators do not show a correlation between local environment and spin parameter, although the slow rotators are more abundant in the highest densities of cluster cores.} For the photometric scaling relations of 
early-type systems
\citet{2009MNRAS.393..798D}  concluded that they are largely independent of the environment  over a large range of mass.

A related, but separate topic is the (stellar population) age and colour of the galaxies,
indicative of the time since the galaxy was quenched. Early-type
dwarfs with blue centres in the Virgo cluster, i.e.~with recent or residual star formation, do not show a clustering towards the cluster centre like other early-type dwarfs \citep{2006AJ....132.2432L}. 
The frequency of such objects increases with decreasing galaxy density of the environment:
\citet{2014MNRAS.445..630P} found  blue centres in the early-type dwarfs
of the Ursa Major cluster to be common.
For the relatively unevolved NGC 5353/4 galaxy group \citet{2008AJ....135.1488T} 
asserted a large fraction of star-forming dwarfs with early-type morphology.
 Likewise, the prototype isolated early-type dwarf
of \citet{2006AJ....131..806G}  has a blue core.
 { Our sample consists by construction (predominantly) of quenched objects.
 This means we will be missing a number of isolated galaxies that have early-type
 morphology but are still forming stars at some level. The fraction of early-type dwarfs with
 blue cores in the Virgo cluster that are excluded by our criteria for quenched galaxies ($\sim$52\%) gives a lower 
 limit for this bias. In this context the study of \citet{2008ApJ...685..904P} is of interest.
 \citeauthor{2008ApJ...685..904P} searched for outliers in the mass (gas-phase) metallicity relation.
 They identified 41 low-mass galaxies ($10^7<M_\star< 10^{10}\, {\rm M}_\odot$) with over-abundant oxygen. These galaxies
 form stars at some level, have early-type morphology, and are typically fairly isolated
 without close companions.  When the nearest-neighbour criteria of our search are employed,
 about half of their objects would be counted as isolated.
 \citet{2008ApJ...685..904P} interpreted the high oxygen abundance
 with a low gas fraction and concluded that their galaxies are transitional dwarfs, evolving from
 star-forming to quenched early-type dwarfs.}

 { The isolated quenched galaxies in our sample (median $M_\star \sim 4\times10^9\, \textrm{M}_\odot$ for the spectroscopic sample)
  suggest that galaxies at the high-mass end of the probed mass range
  can be quenched even without the cluster environment and its ram pressure stripping.
This is consistent with the lower mass limit of \citet[$M_\star< 10^9\, {\rm M}_\odot$]{2012ApJ...757...85G}  
below which no isolated quenched galaxies were found.
We note that our fiducial search criteria for isolated galaxies are less restrictive than 
the analysis of \citet{2012ApJ...757...85G}.  
Their velocity interval for the neighbour search was 1000 km s$^{-1}$, while we use 500 km s$^{-1}$. 
Furthermore, \citet{2012ApJ...757...85G}  found a definition of isolation in terms of the (projected) distance to the massive neighbour: The fraction of quenched galaxies
as a function of the distance to the host flattens out at $\sim$1.5 Mpc. 
This applies for different mass bins and is consistent with our histogram in  Fig.~\ref{fig:dhist}.
}
The lowest mass object of those that we observed, LCBS1123P, has a neighbour
at 1.2 Mpc with $M_{K_s}\sim-23$~mag. We also note that 2MASX~J08192430
has a luminous neighbour at $D_\textrm{proj}=277$ kpc with a velocity difference
of $\Delta V=612$~km~s$^{-1}$.
For the other 8 objects our classification 
does not change when increasing the velocity interval to 1000 km~s$^{-1}$.
This also applies to 28 of the 37 other candidates. 
When counting 4 times less massive galaxies as bright neighbours ($M_{K_s}=-21.5$~mag; the completeness
limit for our bright neighbour search; { corresponding to $M_\star\sim8\times10^9$ M$_\odot$}), the total numbers of candidates reduce
to 17 and 12 for the 500~km~s$^{-1}$ and 1000~km~s$^{-1}$ velocity difference intervals, respectively { (see Table~\ref{table:A1} in the Appendix). 
Even with the most stringent combination for the definition of isolation, i.e.~a neighbour search extending to masses on the same order of magnitude as the low-mass galaxies within a velocity interval of $\pm1000$~km~s$^{-1}$ and separations of at least 1.5 Mpc, 5 of the objects in our spectroscopic sample (and 11 of the complete sample of 46 galaxies) remain classified as isolated.}

Summarizing this section: 
\textit{(i)} our sample of quenched low-mass galaxies in the field does not follow a trend claimed in galaxy clusters
of increasing spin parameter in dwarfs for decreasing local galaxy density; 
\textit{(ii)} our sample consists by construction predominantly
of quenched objects, while early-type dwarfs in low-density environments typically show signs of recent star formation; 
\textit{(iii)} the classification as isolated objects is rather robust when it comes to the velocity interval that is considered for finding
massive neighbours, but the numbers are approximately halved when considering also less massive neighbours; 
and \textit{(iv)} while quenched dwarfs are rare in the field, several examples at the high mass end were identified.

 \subsection{Formation scenarios for isolated early-type dwarfs}
 In principle isolated galaxies need not  have lived in the field all their life \citep[e.g.][]{2015Sci...348..418C}.
 However, the distance to a more massive galaxy in our sample is $D_{\textrm{proj}}\gtrsim1$ Mpc. This distance
 means a travel time of at least 1 Gyr with  a relative velocity of 1000~km~s$^{-1}$.
Such a velocity  exceeds the velocity dispersion of galaxies in the Virgo cluster (e.g.~\citealt{1993A&AS...98..275B}; $\sigma_{\rm Virgo}$$\sim$$700$~km~s$^{-1}$).
 Relative velocities more typical for less massive groups or galaxy pairs increase this 
 hypothetical travel time to the order of a Hubble time.
 
 If the isolated galaxies did not escape a high density environment (or at least from a more massive host galaxy),
 the transformative processes associated with such environments, { e.g.~harassment and ram pressure stripping,}
  cannot be the source for this galaxy population.
 In other words, these galaxies cannot be environmentally transformed spirals, 
 and an alternative  formation process needs to operate.
 
 One possibility is dwarf mergers, which were less frequently considered, since they are unlikely to happen where early-type dwarfs are predominantly found 
 (i.e.~in the cluster environment where large relative velocities make them difficult).  
 However, recently observational evidence has mounted suggesting such mergers
 (occasionally) happen (in other environments) and simulations  have shown  them to be a potential way to also form
 low-mass galaxies \citep{1998AJ....116.1186V,2008MNRAS.388L..10B,2008MNRAS.389.1111V,2012ApJ...748L..24M,2012ApJ...750..121G,2014MNRAS.442.2909C,2015ApJ...812L..14B,2016AJ....151..141P,2016MNRAS.462.3314W}.
  They were also suggested (in the context of pre-processing in galaxy groups) as a possible origin for KDCs in early-type dwarfs \citep{2014ApJ...783..120T}.

Another possibility is that the isolated quenched low-mass galaxies did not experience any dramatic events (such as mergers or harassment) in their past.
Low-mass, pressure supported, 
galaxies  are expected to form as the first galaxies and constitute the building blocks in hierarchical structure formation and galaxy evolution.
A disc can be grown and angular momentum can be obtained  in a more steady fashion by gas accretion (and possibly minor mergers) from the cosmic web \citep[see also][]{2006ApJ...636L..25M}, as also suggested for more massive
early types \citep[and references therein]{2009Natur.457..451D,2015ApJ...804...32G}. 
While this is not required or preferred, such gas accretion  may even be feasible in the 
outskirts of the Virgo cluster as discussed by \citet{2012AJ....144...87H}.

In the context of gas accretion LEDA 218986 is  possibly the most intriguing object in our sample.
 It seems to be a candidate for the addition of gas, which is changing
the internal kinematics, as well as being responsible for the emergence of KDCs (see also \citealt{2013ApJ...770L..26D}) and for the creation of disc features in the image (see Graham et al.~2016 for details). These disc features resemble those in the
Virgo cluster early-type dwarf galaxy VCC~216 (cf.~\citealt{2009AN....330..966L}).

For the objects to be classified as quenched, the star formation activity needs to have ceased.
One way is to cut off the gas supply, which is also consistent with the gas deficiency of the early-type dwarfs 
in clusters.
For galaxies more massive than $M_\star>10^8\, \textrm{M}_\odot $ supernova feedback is not sufficient to permanently expel the gas \citep{2008MNRAS.389.1111V}.
Observationally, \citet{2012ApJ...757...85G} did not find any quenched galaxies in the field with a stellar mass lower than $M_\star<10^9\, \textrm{M}_\odot$.
{ While in clusters the gas supply can be removed by ram pressure stripping, this is not plausible for the most isolated objects.
In principle there could also be gas that just does not form stars.
Or possibly, the gas may have been consumed some time ago for those galaxies in the field that are quenched today,
leading to a `starvation' similar to the scenario (with an external removal of the gas supply) proposed by \citeauthor{1980ApJ...237..692L} (\citeyear{1980ApJ...237..692L}; see also \citealt{2008ApJ...674..742B}).}
 The time scale for the gas consumption  might depend on galaxy mass (like the star formation time-scale, i.e.~downsizing) and  
isolated galaxies with lower mass possibly did not
have enough time to reach this stage.
{  Regardless of the exact mechanisms involved, the  dwarfs observed by
 \citet[see above]{2008ApJ...685..904P} appear to be good candidates for isolated galaxies
with low gas fractions 
  in their final stages before transitioning to being red and dead.
}

In this context it is interesting to note that sometimes blue compact dwarf galaxies (BCDs) are considered as analogs of compact star forming
galaxies typical for earlier epochs of the Universe. It has been suggested that galaxies may evolve through several cycles of a BCD phase, characterised
by intense star formation, and more quiescent phases, in which they resemble more normal low-mass late-type galaxies, to end up
as early-type dwarfs after a final centrally concentrated star formation episode, after which  the gas is consumed \citep{1988MNRAS.233..553D}.
Also, some of the BCDs are observed to have low $v/\sigma$ \citep{2014MNRAS.441..452K}, comparable to the early-type dwarfs { (Fig.~\ref{fig:vsigma}).}
The morphologies and  circular rotation velocity gradient for both galaxy types were also found to compare favourable \citep{Meyer:2014bw,2014A&A...563A..27L}.
Moreover, the stellar populations of  transition type dwarfs appear to differ from those of early-type dwarfs
only in the star formation activity at present day (\citealt{2013MNRAS.428.2949K}, see also \citealt{2008ApJ...685..904P}; and  \citealt{2008MNRAS.385.1374M} for a comparison of stellar population properties of early-type dwarfs in the Virgo cluster and less dense environments).

In summary, we consider both dwarf dwarf mergers and gas accretion as viable candidates for explaining rotating quenched low-mass galaxies in isolation,
bearing in mind also the earlier suggested evolutionary link between early-type dwarfs and BCDs.

\subsection{Implications for early-type dwarfs in clusters}

Our observation of rotating  quenched low-mass  galaxies in isolated environments indicates
 that the cluster environment need not have transformed spirals into early-type galaxies.
 This is  because the  field early-type galaxies have acquired their rotation from their formation 
 as a { less massive} galaxy, rather than from the transformation of { a more massive spiral} galaxy.

The number of early-type dwarfs in the field versus groups versus clusters  \citep[e.g.][]{2002MNRAS.335..712T,2011A&A...528A..19A} 
suggests that the high density environment is still more prolific for forming early-type dwarfs.
Before discussing this within the formation scenarios for isolated quenched low-mass galaxies, 
it should be noted that our observations do not rule out a contribution of { processes related to the high density environments (see \citealt{2009AN....330.1043L}
 for an overview; see also \citealt{2016arXiv160700384M} for a perspective from higher redshift),
such as harassment and ram pressure stripping to the  population of early-type dwarfs in clusters.  The 
latter may be required to quench the galaxies at lower stellar masses. }

The formation scenarios  discussed above can yield more galaxies in high density environments than in the field 
for the following reasons. Firstly, a greater number of galaxies is expected to form in higher over-densities in the early Universe.
Secondly, the high-density environment is likely to speed up the quenching by shutting off the gas supply, i.e.~via ram pressure stripping
\citep[see also][]{2009A&A...498..407G,2011A&A...528A..19A,2013MNRAS.436.1057D}, 
{ and it can also quench  galaxies below the mass limit of \citet{2012ApJ...757...85G}, where there are no quenched galaxies in isolation. 
The galaxies in our sample are (mostly) more massive than those early-type dwarfs in the comparison sample for the Virgo
cluster (the highest stellar mass in that sample is $M_\star \sim1.5\times10^9\,  \textrm{M}_\odot$). 
However, all of the galaxies in our sample can be considered as dwarfs when using the classical $M_B>-18$ mag criterion. 
We also note that the BCDs of \citet[dynamical masses below $10^9$ M$_\odot$]{2014MNRAS.441..452K}  and the objects in \citet[15 of them have $M_\star< 10^9$ M$_\odot$]{2008ApJ...685..904P} fit well into the mass range of the Virgo early-type dwarfs and are in this sense compatible with turning into early-type dwarfs after quenching.}

The  formation scenarios suggested for the quenched low-mass galaxies in isolation are also very attractive as a contributor to the early-type dwarf population in high density environments:
If the progenitor galaxy is a { (more massive)} spiral galaxy,
the transformation has to  alter its morphology, reduce the size and mass,  and remove considerable amounts of angular momentum (\citealt{2015IAUS..311...78F}). 
This { may  even apply for some of the} relatively low-mass late-type galaxies at the end of the Hubble sequence \citep{2014ApJ...789...63A,Janz:2016cw}.
The only feasible way for a substantial transformation seems to be rather strong harassment.
However, \citet{2015MNRAS.454.2502S} showed that this is unlikely to happen for galaxies falling into a cluster \citep[cf. also][]{2013MNRAS.432.1162L,2015A&A...576A.103B}.
The authors put rather conservative limits to the type of orbits which plausibly can lead to strong harassment and concluded
that they are basically only likely for objects that were part of the cluster early on.

If the galaxy, instead of having a spiral galaxy as progenitor, has similar mass, size, and little angular momentum to start with, 
strong harassment is not required. In some cases angular momentum can be obtained by gas  accretion and a disc might be grown.
In other cases the galaxies could remain pressure supported or grow by dwarf dwarf mergers.
Low-mass galaxies in the early Universe, with similar properties to some of the BCDs today (\citealt{Meyer:2014bw,2014MNRAS.441..452K,2015MNRAS.451.1130L}; for simulations of dwarf dwarf mergers forming BCDs and their rotation curves see also \citealt{2016MNRAS.462.3314W}),
appear to be good candidates for evolving into galaxies similar to today's  early-type dwarf galaxies in the galaxy clusters.
{ 
Since there is a mass limit observed below which quenched galaxies are absent in isolation ($M_\star < 10^9$ M$_\odot$; \citealt{2012ApJ...757...85G}; i.e.~in a mass range for which feedback is inefficient),
gas removal via processes like ram pressure stripping appears to be required to quench galaxies  below that mass limit today.}

There is a long-lasting debate whether or not there is a structural dichotomy between
dwarf and ordinary early-type galaxies, i.e.~whether these galaxies follow distinct scaling
relations brighter or fainter than $M_B=-18$~mag (e.g.~\citealt{2009ApJS..182..216K,2012ApJS..198....2K,2013pss6.book...91G}, and many references therein; see also \citealt{1973ApJ...179..423F} and  \citealt{1984ApJ...282...85W}).
{ For example, \citet{2003AJ....125.2936G}  showed that ordinary and dwarf early-type galaxies follow a log-linear relation of profile shape parametrized by the S\'ersic index $n$ (see also \citealt{1997ASPC..116..239J}) with magnitude and demonstrated that this together with a continuous linear relation of central surface brightness and magnitude leads to curved, but continuous  relations involving `effective' or half-light quantities, e.g. in the  size magnitude plane.}
With  rebuttal of the necessity for a different formation scenario (i.e.~galaxy harassment)
for dwarf and ordinary early types, this question can be revisited.
There are no signs for such a dichotomy in the spin parameter (i.e.~when comparing \citealt{Emsellem:2011br} and \citealt{2015ApJ...799..172T}).
{ The   early-type dwarfs  overlap in the spin parameter versus ellipticity diagram with  the ordinary early types, 
but their parameter range is largely separate from that of the spiral galaxies.} 
Furthermore, at intermediate mass, between the dwarf systems and the most
massive galaxies, slow rotators are not frequent.
The frequencies of KDCs in dwarf and ordinary early types are also comparable
\citep{2014ApJ...783..120T}.

We conclude that the { discovery of rotation in quenched low-mass galaxies in isolation}
shows that early-type dwarfs in clusters need not be harassed spirals.
Furthermore, early-type dwarfs that are formed as low-mass galaxies,
{ possibly quenched with the help of the clusters' ram pressure,}
 are an appealing contribution to  the population of cluster dwarfs. 

\section{Summary}
We carried out a search for quenched low-mass galaxies { in low galaxy density environments and isolation} in the Local volume of the SDSS.
The criteria were a stellar mass below $M_\star < 5\times10^9\, \textrm{M}_\odot$, 
a strong 4000 \AA{} break D$_{n}(4000) > 0.6+0.1\times \mathrm{log_{10}}(M_{\star}/\textrm{M}_{\odot})$ and 
an H$\alpha$ equivalent width EW$_{{\rm H}\alpha}< 2$~\AA{},
and no massive neighbour ($M_{\star}\gtrsim3 \times 10^{10}$~M$_{\odot}$) within a velocity interval of 500 km~s$^{-1}$ 
with a distance (in projection) less than   $D_\textrm{proj} < 1$ Mpc.
Our selection criteria yielded 46 galaxies. These were manually checked, e.g.~for star formation missed
by the SDSS fibre, and suitable candidates for spectroscopic follow-up were chosen.
We observed 9 of these objects with Keck ESI and have presented their internal kinematics based on
the obtained spectra.

The nine objects exhibit a large variety of dynamic configurations: some show rotation velocities within
$R_\textrm{e}$ that exceed the random motions, while others have close to no rotation at all.
In one case we  found signs of a kinematically decoupled inner region, which moves aligned with the counter-rotating ionised gas.

We compared our sample of isolated quenched low-mass galaxies to a sample of early-type dwarfs
in the Virgo cluster. { Both samples show similar characteristics: the galaxies have $M_B\gtrsim-18$~mag,}
they mostly lie on the red sequence and span a similar range of sizes. Likewise, their internal kinematics
show a similar variety of behaviour, and the fractions of slow and fast rotations 
are compatible given  the small number statistics.

The {\it rotating} quenched galaxies in our sample cannot be environmentally transformed spiral galaxies due to their isolation.
This has an important implication:  {\it early-type dwarfs in galaxy
clusters need not be transformed spiral galaxies}. 

We discussed dwarf mergers and gas accretion as formation scenarios for our sample of isolated galaxies.
Both of them should be able to produce the required  range of angular momentum and degree of rotational support.
Furthermore, we argued that these processes may contribute a greater number of early-type dwarfs
to high-density environments than the small numbers of them found in the field.
{ And ram pressure stripping can help to account for the quenched galaxies at lower stellar mass.}
Finally, we reasoned that given other challenges to the transformed spiral scenario such a contribution to the 
early-type dwarf population in galaxy clusters is appealing.
The suggested formation scenarios are not exclusive, but their contribution can also help
to explain the heterogeneity of early-type dwarf galaxies.

\section*{Acknowledgements}

We thank the referee for suggestions that helped improve the presentation of our results.
JJ and DAF  thank the ARC for financial support via DP130100388. 
 JJ also thanks T. Lisker for providing the non-parametric photometry code.
SJP acknowledges financial support from the University of Portsmouth.

The NASA-Sloan Atlas was created by Michael Blanton, with extensive help and testing from Eyal Kazin, Guangtun Zhu, Adrian Price-Whelan, John Moustakas, Demitri Muna, Renbin Yan and Benjamin Weaver.

Funding for SDSS-III has been provided by the Alfred P. Sloan Foundation, the Participating Institutions, the National Science Foundation, and the U.S. Department of Energy Office of Science. The SDSS-III web site is \url{http://www.sdss3.org/}.

SDSS-III is managed by the Astrophysical Research Consortium for the Participating Institutions of the SDSS-III Collaboration including the University of Arizona, the Brazilian Participation Group, Brookhaven National Laboratory, Carnegie Mellon University, University of Florida, the French Participation Group, the German Participation Group, Harvard University, the Instituto de Astrofisica de Canarias, the Michigan State/Notre Dame/JINA Participation Group, Johns Hopkins University, Lawrence Berkeley National Laboratory, Max Planck Institute for Astrophysics, Max Planck Institute for Extraterrestrial Physics, New Mexico State University, New York University, Ohio State University, Pennsylvania State University, University of Portsmouth, Princeton University, the Spanish Participation Group, University of Tokyo, University of Utah, Vanderbilt University, University of Virginia, University of Washington, and Yale University.

We acknowledge the usage of the HyperLeda database (\citealt{2014A&A...570A..13M}; http://leda.univ-lyon1.fr).
This research has made use of the NASA/IPAC Extragalactic Database (NED) which is operated by the Jet Propulsion Laboratory, California Institute of Technology, under contract with the National Aeronautics and Space Administration.
This research has made use of the SIMBAD database,
operated at CDS, Strasbourg, France 

This publication makes use of data products from the Two
Micron All Sky Survey, which is a joint project of the University of
Massachusetts and the Infrared Processing and Analysis
Center/California Institute of Technology, funded by the National
Aeronautics and Space Administration and the National Science
Foundation.



\bibliographystyle{mnras}



\appendix
\section{Alternative neighbour search}
Here we include the parameters of the nearest bright neighbour when searching for neighbours with $M_{K_s}<-21.5$ mag (Table~\ref{table:A1}).
\begin{table}
\begin{center}
\caption{Projected distances and velocity differences for luminous neighbours with $M_{K_s}<-21.5$ mag. \label{table:A1}}
\begin{tabular}{lcccc}
\hline \hline
Galaxy         &  \multicolumn{4}{c}{closest luminous neighbour}\\
 & $D_{500}$ & $|\Delta V_{500}|$ & $D_{1000}$ & $|\Delta V_{1000}|$  \\
 	&       [Mpc] &  [km s$^{-1}$]  &  [Mpc] & [km s$^{-1}$]  \\
\hline
LEDA~3115955  &  1.67 &   13 \\
2MASX~J03190758+4232179 &  0.63 &  311  &  0.55 &  586 \\
LCSBS1123P  &  1.24 &  167 \\
2MASX~J08192430+2100125 &  0.20 &  203  &  0.06 &  950 \\
VIIIZw040  &  3.47 &   62 \\
CGCG038-085  &  2.03 &   75 \\
2MASX~J11521124+0421239 &  3.14 &  166  &  1.38 &  662 \\
CGCG101-026  &  2.40 &  263 \\
LEDA~2108986  &  3.07 &   21 \\
\hline
J001212.41--110010.4  &  0.10 &   29 \\
J001530.03+160429.7  &  0.21 &  384 \\
J001601.19+160133.4  &  0.34 &  214 \\
J012506.69--000807.0  &  0.67 &  293 \\
J013842.89--002053.0  &  0.53 &   80 \\
J045058.77+261313.7 &  1.90 &  305  &  1.83 &  661 \\
J075303.96+524435.8  &  1.04 &   63 \\
J082013.92+302503.0  &  0.88 &   97 \\
J082210.66+210507.5  &  0.16 &  482 \\
J084915.01+191127.3 &  0.22 &  500  &  0.11 &  526 \\
J085652.63+475923.8  &  3.67 &  277 \\
J091514.45+581200.3  &  1.53 &   58 \\
J091657.98+064254.2  &  0.77 &   66 \\
J093016.38+233727.9  &  0.04 &   52 \\
J093251.11+314145.0  &  1.72 &  209 \\
J094408.52+111514.8  &  0.08 &   84 \\
J094834.50+145356.6  &  0.46 &   94 \\
J100003.93+044845.0  &  0.19 &   66 \\
J105005.53+655015.6  &  0.10 &  162 \\
J110423.34+195501.5  &  0.56 &  263 \\
J112422.89+385833.3 &  1.30 &  354  &  0.06 &  925 \\
J114423.13+163304.5 &  1.21 &  331  &  0.02 &  549 \\
J120300.94+025011.0  &  1.06 &   55 \\
J120823.99+435212.4  &  0.77 &   27 \\
J122543.23+042505.4  &  0.11 &  209 \\
J124408.62+252458.2  &  0.14 &    2 \\
J125026.61+264407.1  &  0.04 &  200 \\
J125103.34+262644.6  &  0.44 &  373 \\
J125321.68+262141.1  &  0.22 &  360 \\
J125756.52+272256.2  &  0.26 &  369 \\
J125940.10+275117.7 &  4.20 &  123  &  0.04 &  936 \\
J130320.35+175909.7 &  1.04 &  164  &  0.96 &  553 \\
J130549.09+262551.6  &  0.91 &  207 \\
J142914.46+444156.3  &  0.89 &   82 \\
J144621.10+342214.1  &  1.97 &  379 \\
J160810.70+313055.0  &  5.52 &   72 \\
J232028.21+150420.8 &  1.02 &  438  &  0.84 &  559 \\
\hline
\end{tabular}
\end{center}
\parbox{0.5\textwidth}{{\bf Notes:   }  
The first column lists the galaxy name.  The closest bright neighbour within a velocity interval of $\pm 500$ km~s$^{-1}$ is extracted from a combined catalogue containing redshifts from the NASA Sloan Atlas and the 2MASS redshift survey.  Instead of $M_{K_s}<-23$ mag as in Table~\ref{table:specdEs}, $M_{K_s}<-21.5$ mag is used.
The projected (linear) distance between the neighbour and the low-mass galaxy and the difference between their recession velocities are given in columns 2 and 3.
If there is an additional neighbour within a velocity interval of $\pm 1000$ km~s$^{-1}$, the corresponding details are listed in columns 4 and 5.} 
\end{table}

\bsp	
\label{lastpage}
\end{document}